\documentstyle[12pt]{report}
\begin{document}
\begin{centering}
\large\bf A complementary group technique for the 
resolution of
the \\
outer multiplicity problem of SU(n).
(II) A recoupling
 approach \\to the solution of 
$SU(3)\supset U(2)$ reduced Wigner coefficients
\vskip .6truecm
\normalsize Feng Pan$^{\dagger}$ and J. P. Draayer
\vskip .2cm
\noindent{\small\it Department of Physics {\normalsize\&} Astronomy, Louisiana State University,}\\
{\small\it Baton Rouge, LA 70803-4001}\\
\vskip 1truecm
{\bf Abstract}\\
\end{centering}
\vskip .5cm
   \normalsize  A general procedure for the derivation of $SU(3)\supset U(2)$ reduced Wigner coefficients
for the coupling $(\lambda_{1}\mu_{1})\times (\lambda_{2}\mu_{2})\downarrow (\lambda\mu)^{\eta}$, where
$\eta$ is the outer multiplicity label needed in the decomposition, is proposed based on a recoupling approach
according to the complementary group technique given in (I). It is proved that the non-multiplicity-free
reduced Wigner coefficients of $SU(n)$ are not unique with respect to canonical outer multiplicity labels,
and can be transformed from one set of outer multiplicity labels to another. The transformation matrices are
elements of $SO(m)$, where $m$ is the number of occurrence of the corresponding irrep $(\lambda\mu)$ in the
decomposition
$(\lambda_{1}\mu_{1})\times (\lambda_{2}\mu_{2})\downarrow (\lambda\mu)$. Thus, a kind of the reduced Wigner
coefficients with multiplicity is obtained after a special $SO(m)$ transformation. New features of this
kind of reduced Wigner coefficients and the differences from the
reduced Wigner coefficients with other choice of the multiplicity label given previously are discussed. The
method can also be applied to the derivation of general $SU(n)$ Wigner or reduced Wigner coefficients with
multiplicity. Algebraic expression of another kind of reduced Wigner coefficients, 
the so-called reduced auxiliary 
Wigner coefficients for $SU(3)\supset U(2)$, are also obtained.
\vskip 1cm

\vskip 4.5cm
\noindent {-----------------------------------------}\\
\noindent $^{\dagger}$On leave from Department of Physics,
Liaoning Normal Univ., Dalian 116029, P. R.~~China

\newpage

\begin{centering}
{\large I. Introduction}\\
\end{centering}
\vskip .4truecm
    Wigner  Coefficients (WCs), or  Reduced Wigner Coefficients (RWCs) of $SU(3)$ in the canonical basis, i.e.
the basis adapted to $SU(3)\supset U(2)$, were discussed by many authors, for example, by Biedenharn et
al$^{[1-10]}$ using the canonical unit tensor operator method, Moshinsky et al$^{[11-14]}$ using the
infinitesimal approach and a complementary group method, Ali\v{s}auskas et al using the symmetric group
approach$^{[15-17]}$ and paracanonical and pseudo-canonical coupling schemes.$^{[18-21]}$ A large class of the
RWCs were also  considered by Hecht,$^{[22]}$ Resnikoff,$^{[23]}$ Shelepin and Karasev,$^{[24-25]}$ Klimyk and
Gavrilik,$^{[26]}$ Le Blanc and Rowe,$^{[27]}$  and many others. Among these approaches, only the outer
multiplicity labeling scheme of the unit tensor operator method is canonical, which leads to the usual
orthogonalities of the RWCs. Other methods for labeling the outer multiplicity are noncanonical, i. e.
the RWCs obtained will be non-orthogonal with respect to the outer multiplicity label. Therefore, the
Gram-Schmidt process will be adopted, which depends upon an arbitrary choice of order to the elements
to be orthogonalized. In this case, only numerical algorithm is possible, as have been done so by 
Draayer and Akiyama,$^{[28-29]}$ and Kaeding and Williams.$^{[30-32]}$  In [33],
WRCs associated to 27-plet operator, which is a multiplicity three case, were discussed.
\vskip .3cm
   Very recently, Parkash and Sharatchandra$^{[34]}$ have worked out an algebraic formula for the general
Wigner coefficients for $SU(3)$ in the canonical basis. However, the final results are expressed in terms of
free summations over 33 variables under some restrictions, and there is a normalization factor needed to be
determined in the expression, which will not be easy to compute values of the RWCs by using their formula
either algebraically or numerically.
\vskip .3cm
   In this paper, we will use the complementary group  technique proposed in (I) to compute the RWCs of
$SU(3)\supset U(2)$ by using the multiplicity-free RWCs known previously with a recoupling approach.
It should be noted that this approach, in principle, is labeling scheme independent. For example, the RWCs
of $SU(3)\supset U(2)$ in another labeling scheme proposed by Biedenhran et al can also be derived by using 
this method. However, the values of the RWCs with different canonical outer multiplicity  labeling scheme will
be different, which are actually within a $SO(m)$ group transformation among the RWCs from one set of outer
multiplicity label to another.
\vskip .3cm
   In Sec. II, we will discuss the non-uniqueness of the canonical resolution of RWCs with multiplicity, and
prove that the transformation group from one set of multiplicity label to another can be chosen  as $SO(m)$,
where
$m$ is the number of occurrence of the resultant irrep considered in the decomposition $(\lambda_{1}\mu_{1})
\times (\lambda_{2}\mu_{2})$. Then, a special $SO(m)$ transformation is made for the RWCs. In such a case,
the RWCs with multiplicity can be derived recursively. In Sec. III, we will propose a recoupling procedure
for the evaluation of the RWCs from known multiplicity-free RWCs of $SU(n)\supset U(n-1)$ given previously
by Ali\v{s}auskas et al.$^{[15]}$  New feature of this kind of RWCs and differences from ones
with other choice of the outer multiplicity label will be discussed in IV. In Sec. V, an analytical expression
of another kind of RWCs, the so-called reduced auxiliary WCs proposed by Brody, Moshinsky, and Renero$^{[11]}$
will also be derived by using this recoupling approach. Discussions  will be given in VI.

\vskip .3cm
\begin{centering}
{\large II. Relations among different choices of the multiplicity labels  }\\
\end{centering}
\vskip .4cm
   Let $\vert [\lambda ]\xi,\rho>\equiv\vert [\lambda_{1}][\lambda_{2}][\lambda ](\xi),\rho>$ be coupled
basis vectors of $U(n)\times U(n)\downarrow U(n)$, where $\xi$ denotes a set of multiplicity labels needed
in the decomposition $[\lambda_{1}]\times [\lambda_{2}]\downarrow [\lambda ]$, $\rho$ is the sublabels
for the resultant irrep $[\lambda ]$. The completeness condition for $\vert [\lambda ]\xi,\rho>$ is
\vskip .3cm
$$\sum_{(\xi)[\lambda ]\rho} \vert [\lambda ]\xi,\rho><[\lambda ]\xi,\rho\vert~=~1.\eqno(2.1)$$
\vskip .3cm
    Actually, the choice of the multiplicity labels is not unique. One can use transformations from
one set of multiplicity labels to another. If $(\xi)$ can be taken $m$ different values denoted as
$(\xi)=\xi_{1},~\xi_{2},~\cdots,~\xi_{m}$, the transformation matrix group is $SU(m)$. Therefore,
\vskip .3cm
$$\vert [\lambda ]\eta,\rho>=\sum_{(\xi)}y(\xi,\eta)\vert [\lambda ]\xi,~\rho>,\eqno(2.2)$$
\vskip .3cm
\noindent where $(\eta)$ is another set of multiplicity labels for $U(n)$, $y(\xi,\eta)$ is a
matrix element of $SU(m)$. One can verify that $y(\xi,\eta)$ should satisfy
\vskip .3cm
$$\sum_{(\xi)}y(\xi,\eta)y^{*}(\xi,\eta^{\prime})=\delta_{\eta\eta^{\prime}},$$

$$\sum_{(\eta)}y(\xi,\eta)y^{*}(\xi^{\prime},\eta)=\delta_{\xi\xi^{\prime}}.\eqno(2.3)$$
\vskip .3cm
\noindent Therefore, $\{y(\xi,\eta)\}$ defines a unitary transformation according to the basic representation
of $SU(m)$. Hence, the choice of the outer multiplicity labels for the Kronecker product $[\lambda_{1}]\times
[\lambda_{2}]\downarrow [\lambda ]$ is not unique. There always exists a unitary transformation ${\bf Y}\in
SU(m)$, where $m$ is the number of occurrence of $[\lambda ]$ in the decomposition $[\lambda_{1}]\times
[\lambda_{2}]$, which transforms from one set of outer multiplicity labels to another. Usually, the WCs of $U(n)$
are taken to be real. In this case, the internal symmetry group for the transformation of the outer multiplicity
labels can be chosen as $SO(m)$. It is obvious that the dimension of the transformation group is the number of
occurrence of the resultant irrep $[\lambda ]$ dependent on the decomposition $[\lambda_{1}]\times
[\lambda_{2}]\downarrow [\lambda ]$, and the basis vectors of $[\lambda ]$ are still orthonormal with each
other if they are transformed from  a set of orthonormal basis vectors of $[\lambda ]$. That is why there are
different forms of RWCs or WCs with respect to the outer multiplicity labels after a Gram-Schmidt
transformation from noncanonical resolutions or obtained from canonical resolutions. In the following, we
always assume that the RWCs considered are real.
\vskip .3cm
  We restrict our discussion to $SU(3)\supset U(2)$ case only, which can easily be
extended to the general $SU(n)$ case.
We adopt the usual physical notation for $SU(3)$ irrep $(\lambda\mu)\equiv [\lambda +\mu,\mu ]$, where 
$[\lambda+\mu,\mu ]$ is a usual two-rowed irrep corresponding to the irrep described by two-rowed Young diagram
with
$\lambda+\mu$ boxes in the first row, and $\mu$ boxes in the second row. We have proved that there is only one
multiplicity label needed in the decomposition
$(\lambda_{1}\mu_{1})\times(\lambda_{2}\mu_{2})\downarrow [m_{1}m_{2}m_{3}]$, which is assigned as $\xi$ with
$\xi=\xi_{1},~\xi_{2},~\cdots,~\xi_{m}$, where $m$ is the number of occurrence of $[m_{1}m_{2}m_{3}]$ in
the decomposition.
\vskip .3cm
   The key step to evaluate $SU(3)$ WCs or RWCs of $SU(3)\supset U(2)$ is to use the transforamtion
\vskip .3cm
$$\sum_{\xi}\left<
\begin{array}{ll}
(\lambda_{1}\mu_{1}) &(\lambda_{2}\mu_{2})\\
~~~\rho_{1} &~~~\rho_{2}
\end{array}\right\vert\left.
\begin{array}{l}
\xi~[m_{1}m_{2}m_{3}]\\
~~~~~~~~\rho
\end{array}\right>y(\xi,\eta)=\left<
\begin{array}{ll}
(\lambda_{1}\mu_{1}) &(\lambda_{2}\mu_{2})\\
~~~\rho_{1} &~~~\rho_{2}
\end{array}\right\vert\left.
\begin{array}{l}
\eta~[m_{1}m_{2}m_{3}]\\
~~~~~~~~\rho
\end{array}\right>,\eqno(2.4)$$
\vskip .3cm
\noindent where $\rho_{1},~\rho_{2}$, and $\rho$ are the corresponding sublabels of $SU(3)$ if 
\vskip .3cm
$$\left<
\begin{array}{ll}
(\lambda_{1}\mu_{1}) &(\lambda_{2}\mu_{2})\\
~~~\rho_{1} &~~~\rho_{2}
\end{array}\right\vert\left.
\begin{array}{l}
\xi~[m_{1}m_{2}m_{3}]\\
~~~~~~~~\rho
\end{array}\right>$$
\vskip .3cm
\noindent is WCs, or RWCs if $\rho_{1},~\rho_{2},~\rho$ are the corresponding U(2) labels, and $y(\xi,\eta)$ are
chosen to be a special set of matrix elements of the special orthogonal group $SO(m)$, which are chosen as follows.
\vskip .3cm
   Assume 
\vskip .3cm
$$\left<
\begin{array}{ll}
(\lambda_{1}\mu_{1}) &(\lambda_{2}\mu_{2})\\
~~~\rho_{1} &~~~\rho_{2}
\end{array}\right\vert\left.
\begin{array}{l}
\xi~[m_{1}m_{2}m_{3}]\\
~~~~~~~~\rho
\end{array}\right>,~~\xi=\xi_{1},~\xi_{2},~\cdots,~\xi_{m},\eqno(2.5)$$
\vskip .3cm
\noindent  is a set of WCs satisfying the orthogonality relation
\vskip .3cm
$$\sum_{\rho_{1}\rho_{2}}\left<
\begin{array}{ll}
(\lambda_{1}\mu_{1}) &(\lambda_{2}\mu_{2})\\
~~~\rho_{1} &~~~\rho_{2}
\end{array}\right\vert\left.
\begin{array}{l}
\xi~[m_{1}m_{2}m_{3}]\\
~~~~~~~~\rho
\end{array}\right>\left<
\begin{array}{ll}
(\lambda_{1}\mu_{1}) &(\lambda_{2}\mu_{2})\\
~~~\rho_{1} &~~~\rho_{2}
\end{array}\right\vert\left.
\begin{array}{l}
\xi^{\prime}~[m^{\prime}_{1}m^{\prime}_{2}m^{\prime}_{3}]\\
~~~~~~~~\rho^{\prime}
\end{array}\right>=\delta_{m_{i}m^{\prime}_{i}}\delta_{\xi\xi^{\prime}}
\delta_{\rho\rho^{\prime}},\eqno(2.6a)$$
\vskip .3cm
$$\sum_{\xi\rho m_{i}}\left<
\begin{array}{ll}
(\lambda_{1}\mu_{1}) &(\lambda_{2}\mu_{2})\\
~~~\rho_{1} &~~~\rho_{2}
\end{array}\right\vert\left.
\begin{array}{l}
\xi~[m_{1}m_{2}m_{3}]\\
~~~~~~~~\rho
\end{array}\right>\left<
\begin{array}{ll}
(\lambda_{1}\mu_{1}) &(\lambda_{2}\mu_{2})\\
~~~\rho^{\prime}_{1} &~~~\rho^{\prime}_{2}
\end{array}\right\vert\left.
\begin{array}{l}
\xi~[m_{1}m_{2}m_{3}]\\
~~~~~~~~\rho
\end{array}\right>=\delta_{\rho_{1}\rho_{1}^{\prime}}\delta_{\rho_{2}\rho_{2}^{\prime}}
,\eqno(2.6b)$$
\vskip .3cm
\noindent or RWCs satisfying 
$$\sum_{\xi m_{i}}\left<
\begin{array}{ll}
(\lambda_{1}\mu_{1}) &(\lambda_{2}\mu_{2})\\
~~~\rho_{1} &~~~\rho_{2}
\end{array}\right\vert\left.
\begin{array}{l}
\xi~[m_{1}m_{2}m_{3}]\\
~~~~~~~~\rho
\end{array}\right>\left<
\begin{array}{ll}
(\lambda_{1}\mu_{1}) &(\lambda_{2}\mu_{2})\\
~~~\rho^{\prime}_{1} &~~~\rho^{\prime}_{2}
\end{array}\right\vert\left.
\begin{array}{l}
\xi~[m_{1}m_{2}m_{3}]\\
~~~~~~~~\rho
\end{array}\right>=\delta_{\rho_{1}\rho_{1}^{\prime}}\delta_{\rho_{2}\rho_{2}^{\prime}}.\eqno(2.6c)$$
\vskip .3cm
$$\sum_{\rho_{1}\rho_{2}}\left<
\begin{array}{ll}
(\lambda_{1}\mu_{1}) &(\lambda_{2}\mu_{2})\\
~~~\rho_{1} &~~~\rho_{2}
\end{array}\right\vert\left.
\begin{array}{l}
\xi~[m_{1}m_{2}m_{3}]\\
~~~~~~~~\rho
\end{array}\right>\left<
\begin{array}{ll}
(\lambda_{1}\mu_{1}) &(\lambda_{2}\mu_{2})\\
~~~\rho_{1} &~~~\rho_{2}
\end{array}\right\vert\left.
\begin{array}{l}
\xi^{\prime}~[m^{\prime}_{1}m^{\prime}_{2}m^{\prime}_{3}]\\
~~~~~~~~\rho
\end{array}\right>=\delta_{m_{i}m^{\prime}_{i}}\delta_{\xi\xi^{\prime}},\eqno(2.6d)$$
\vskip .3cm
\noindent According to the Schur-Weyl duality relation, the RWCs for $SU(3)\supset U(2)$ given by (2.5) are also
RWCs for $SU(4)\supset U(3)$ for the same coupling with the same set of outer multiplicity labels $\{\xi_{i}\}$.
Then, according to the complementary group technique given by (I), the complementary group for the above $SU(3)$
coupling is ${\cal U}(4)$. We need to consider the same coupling of ${\cal U}(4)$  in the special Gel'fand basis
according to the Littlewood rule, namely
\vskip .3cm
$$\left<
\matrix{ {[\lambda_{1}+\mu_{1},~\mu_{1}]} &[\lambda_{2}+\mu_{2},~\mu_{2}]\cr
{[\lambda_{1}+\mu_{1},~\mu_{1}]} &[\lambda_{2}+\mu_{2},~0]\cr
}
\right\vert\left.
\matrix{
\xi~[m_{1}m_{2}m_{3}]\cr
[m_{1},~m_{2}-\eta,~m_{3}-\mu_{2}+\eta ]\cr
}\right>.\eqno(2.7)$$
\vskip .3cm
\noindent Next, there is a $m\times m$ orthonormal matrix $Y$ that transforms the RWCs or WCs between two sets
of multiplicity labels $(\xi)$ and $(\tilde{\eta})$, where the range of $\xi$, and $\tilde{\eta}$ is the same as
the sublabels
$\eta$ in the $SU(3)$ subirrep $[m_{1},~m_{2}-\eta,~m_{3}-\mu_{2}+\eta ]$. It has been proved  that the number
of occurrence of $[m_{1},m_{2},m_{3}]$ in the Kronecker product $[\lambda_{1}+\mu_{1},\mu_{1}]\times
[\lambda_{2}+\mu_{2},\mu_{2}]$ can be described exactly by $\eta$ within the following ranges
\vskip .3cm
$$\eta_{\min}\leq\eta\leq\eta_{\max},\eqno(2.8a)$$
\vskip .3cm
\noindent where
\vskip .3cm
$$\eta_{\max}=\min(m_{1}-\lambda_{1}-\mu_{1},~\mu_{2},~m_{2}-\mu_{1},~\lambda_{2}+\mu_{2}-m_{3},~
\mu_{1}+\mu_{2}-m_{3},~m_{2}-m_{3}),$$

$$\eta_{\min}=\max( 0,~\mu_{2}-m_{3},~m_{2}-\lambda_{1}-\mu_{1}).\eqno(2.8b)$$
\vskip .3cm 
\noindent Thus, we require
\vskip .3cm
$$\sum_{\xi}y(\xi,\tilde{\eta})\left<
\matrix{
[\lambda_{1}+\mu_{1},~\mu_{1}] &[\lambda_{2}+\mu_{2},~\mu_{2}]\cr
[\lambda_{1}+\mu_{1},~\mu_{1}] &[\lambda_{2}+\mu_{2},~0]\cr
}
\right\vert\left.
\matrix{
\xi~[m_{1}m_{2}m_{3}]\cr
[m_{1},~m_{2}-\eta,~m_{3}-\mu_{2}+\eta ]\cr}
\right>$$

$$=\left<
\matrix{
[\lambda_{1}+\mu_{1},~\mu_{1}] &[\lambda_{2}+\mu_{2},~\mu_{2}]\cr
[\lambda_{1}+\mu_{1},~\mu_{1}] &[\lambda_{2}+\mu_{2},~0]\cr
}
\right\vert\left.
\matrix{
\tilde{\eta}~[m_{1}m_{2}m_{3}]\cr
[m_{1},~m_{2}-\eta,~m_{3}-\mu_{2}+\eta ]\cr}
\right>^{\prime},\eqno(2.9)$$
\vskip .3cm
\noindent  where $\tilde{\eta},~\eta~=\eta_{1},~\eta_{2},\cdots,~\eta_{m}$, and the prime indicates 
a new RWC, which is different from the old one. The RWCs 
\vskip .3cm
$$\left<{\xi\over{\eta}}\right>\equiv\left<
\matrix{
[\lambda_{1}+\mu_{1},~\mu_{1}] &[\lambda_{2}+\mu_{2},~\mu_{2}]\cr
[\lambda_{1}+\mu_{1},~\mu_{1}] &[\lambda_{2}+\mu_{2},~0]\cr
}
\right\vert\left.
\matrix{
\xi~[m_{1}m_{2}m_{3}]\cr
[m_{1},~m_{2}-\eta,~m_{3}-\mu_{2}+\eta ]\cr
}\right>\eqno(2.10)$$
\vskip .3cm
\noindent  with fixed $\eta$ can be regarded as a vector in ${\bf R}^{m}$ space.
\vskip .3cm
$$\left<{\xi\over{\eta}}\right>,~~\xi=\xi_{1},~\xi_{2},\cdots,~\xi_{m}.\eqno(2.11)$$
\vskip .3cm
 We choose the following special transformation such that
\vskip .3cm
$${\bf Y}\left(
\matrix{
\left<{\eta_{1}\over{\eta_{1}}}\right>\cr
{}\cr
\left<{\eta_{2}\over{\eta_{1}}}\right>\cr
{}\cr
\vdots\cr
{}\cr
\left<{\eta_{m}\over{\eta_{1}}}\right>\cr}
\right)=\left(
\matrix{
\left<{\eta_{1}\over{\eta_{1}}}\right>^{\prime}\cr
{}\cr
0\cr
{}\cr
\vdots\cr
{}\cr
0\cr}
\right).\eqno(2.12)$$
\vskip .3cm
\noindent Therefore, all the components of the old vectors $\left<{\xi\over{\eta_{1}}}\right>$
can be expressed by the following relation
\vskip .3cm
$$\left<{\xi\over{\eta_{1}}}\right>=y(\eta_{1},\xi)\left<{\eta_{1}\over{\eta_{1}}}\right>.\eqno(2.13)$$
\vskip .3cm
\noindent  While other $m-1$ vectors $\left<{\xi\over{\eta_{i}}}\right>,~i=2,~3,\cdots,~m$, also
undergo the same transformation. We choose
\vskip .3cm
$${\bf Y}\left(
\matrix{
\left<{\eta_{1}\over{\eta_{2}}}\right>\cr
{}\cr
\left<{\eta_{2}\over{\eta_{2}}}\right>\cr
{}\cr
\vdots\cr
{}\cr
\left<{\eta_{m}\over{\eta_{2}}}\right>\cr}
\right)=\left(
\matrix{
\left<{\eta_{1}\over{\eta_{2}}}\right>^{\prime}\cr
{}\cr
\left<{\eta_{2}\over{\eta_{2}}}\right>^{\prime}\cr
{}\cr
0\cr
\vdots\cr
0\cr}
\right),~~{\bf Y}\left(
\matrix{
\left<{\eta_{1}\over{\eta_{3}}}\right>\cr
{}\cr
\left<{\eta_{2}\over{\eta_{3}}}\right>\cr
{}\cr
\left<{\eta_{3}\over{\eta_{3}}}\right>\cr
\vdots\cr
{}\cr
\left<{\eta_{m}\over{\eta_{3}}}\right>\cr}
\right)=\left(
\matrix{
\left<{\eta_{1}\over{\eta_{3}}}\right>^{\prime}\cr
{}\cr
\left<{\eta_{2}\over{\eta_{3}}}\right>^{\prime}\cr
{}\cr
\left<{\eta_{3}\over{\eta_{3}}}\right>^{\prime}\cr
{}\cr
0\cr
\vdots\cr
0\cr}
\right),~~\cdots\cdots,$$
\vskip .3cm
$${\bf Y}\left(
\matrix{
\left<{\eta_{1}\over{\eta_{m-1}}}\right>\cr
{}\cr
\left<{\eta_{2}\over{\eta_{m-1}}}\right>\cr
{}\cr
\vdots\cr
{}\cr
\left<{\eta_{m}\over{\eta_{m-1}}}\right>\cr}
\right)=\left(
\matrix{
\left<{\eta_{1}\over{\eta_{m-1}}}\right>^{\prime}\cr
{}\cr
\left<{\eta_{2}\over{\eta_{m-1}}}\right>^{\prime}\cr
\vdots\cr
\left<{\eta_{m-1}\over{\eta_{m-1}}}\right>^{\prime}\cr
{}\cr
0\cr
}
\right),~~{\bf Y}\left(
\matrix{
\left<{\eta_{1}\over{\eta_{m}}}\right>\cr
{}\cr
\left<{\eta_{2}\over{\eta_{m}}}\right>\cr
{}\cr
\vdots\cr
{}\cr
\left<{\eta_{m}\over{\eta_{m}}}\right>\cr}
\right)=\left(
\matrix{
\left<{\eta_{1}\over{\eta_{m}}}\right>^{\prime}\cr
{}\cr
\left<{\eta_{2}\over{\eta_{m}}}\right>^{\prime}\cr
{}\cr
\vdots\cr
{}\cr
\left<{\eta_{m}\over{\eta_{m}}}\right>^{\prime}\cr
}
\right),\eqno(2.14)$$
\vskip .3cm
\noindent  where the zero components (the  special RWCs) have clearly been written out after the transformation.
Other components are non-zero in general.
\vskip .3cm
   In the following, we give an example for $m=3$ case to show such transformation is always possible. In $m=3$
case, one can make the following special rotation $A_{1}$ around the third axis such that
\vskip .3cm
$$\left(\matrix{
cos\theta_{1} &-sin\theta_{1} &{}\cr
{}\cr
sin\theta_{1} &~cos\theta_{1} &{}\cr
{}\cr
{} &{} &1\cr}\right)\left(\matrix{
\left<{\eta_{1}\over{\eta_{1}}}\right>\cr
\cr
\left<{\eta_{2}\over{\eta_{1}}}\right>\cr
\cr
\left<{\eta_{3}\over{\eta_{1}}}\right>\cr}\right)=\left(\matrix{
\left<{\eta_{1}\over{\eta_{1}}}\right>^{\prime}\cr
\cr
0\cr
\cr
\left<{\eta_{3}\over{\eta_{1}}}\right>\cr}\right),
\left(\matrix{
cos\theta_{1} &-sin\theta_{1} &{}\cr
{}\cr
sin\theta_{1} &~cos\theta_{1} &{}\cr
{}\cr
{} &{} &1\cr}\right)\left(\matrix{
\left<{\eta_{1}\over{\eta_{2}}}\right>\cr
\cr
\left<{\eta_{2}\over{\eta_{2}}}\right>\cr
\cr
\left<{\eta_{3}\over{\eta_{2}}}\right>\cr}\right)=\left(\matrix{
\left<{\eta_{1}\over{\eta_{2}}}\right>^{\prime}\cr
\cr
\left<{\eta_{2}\over{\eta_{2}}}\right>^{\prime}\cr 
\cr
\left<{\eta_{3}\over{\eta_{2}}}\right>^{\prime}\cr}\right),$$

$$\left(\matrix{
cos\theta_{1} &-sin\theta_{1} &{}\cr
{}\cr
sin\theta_{1} &~cos\theta_{1} &{}\cr
{}\cr
{} &{} &1\cr}\right)\left(\matrix{
\left<{\eta_{1}\over{\eta_{3}}}\right>\cr
\cr
\left<{\eta_{2}\over{\eta_{3}}}\right>\cr
\cr
\left<{\eta_{3}\over{\eta_{3}}}\right>\cr}\right)=\left(\matrix{
\left<{\eta_{1}\over{\eta_{3}}}\right>^{\prime}\cr
\cr
\left<{\eta_{2}\over{\eta_{3}}}\right>^{\prime}\cr 
\cr
\left<{\eta_{3}\over{\eta_{3}}}\right>^{\prime}\cr}\right),\eqno(2.15)$$
\vskip .3cm
\noindent Then, make another rotation $A_{2}$ around the second axis with
\vskip .3cm
$$\left(\matrix{
cos\theta_{2} &{} &-sin\theta_{2} \cr
{}\cr
0 &1 &0\cr
\cr
sin\theta_{2} &{} &~cos\theta_{2} \cr
}\right)\left(\matrix{
\left<{\eta_{1}\over{\eta_{1}}}\right>^{\prime}\cr
\cr
0\cr
\cr
\left<{\eta_{3}\over{\eta_{1}}}\right>\cr
}\right)=\left(\matrix{
\left<{\eta_{1}\over{\eta_{1}}}\right>^{\prime\prime}\cr
\cr
0\cr
\cr
0\cr}\right),
\left(\matrix{
cos\theta_{2} &{} &-sin\theta_{2} \cr
{}\cr
0 &1 &0\cr
\cr
sin\theta_{2} &{} &~cos\theta_{2} \cr
}\right)\left(\matrix{
\left<{\eta_{1}\over{\eta_{2}}}\right>^{\prime}\cr
\cr
\left<{\eta_{2}\over{\eta_{2}}}\right>^{\prime}\cr
\cr
\left<{\eta_{3}\over{\eta_{2}}}\right>^{\prime}\cr
}\right)=\left(\matrix{
\left<{\eta_{1}\over{\eta_{2}}}\right>^{\prime\prime}\cr
\cr
\left<{\eta_{2}\over{\eta_{2}}}\right>^{\prime\prime}\cr
\cr
\left<{\eta_{2}\over{\eta_{2}}}\right>^{\prime\prime}\cr
}\right),$$

$$\left(\matrix{
cos\theta_{2} &{} &-sin\theta_{2} \cr
{}\cr
0 &1 &0\cr
\cr
sin\theta_{2} &{} &~cos\theta_{2} \cr
}\right)\left(\matrix{
\left<{\eta_{1}\over{\eta_{3}}}\right>^{\prime}\cr
\cr
\left<{\eta_{2}\over{\eta_{3}}}\right>^{\prime}\cr
\cr
\left<{\eta_{3}\over{\eta_{3}}}\right>^{\prime}\cr
}\right)=\left(\matrix{
\left<{\eta_{1}\over{\eta_{3}}}\right>^{\prime\prime}\cr
\cr
\left<{\eta_{2}\over{\eta_{3}}}\right>^{\prime\prime}\cr
\cr
\left<{\eta_{2}\over{\eta_{3}}}\right>^{\prime\prime}\cr
}\right),\eqno(2.16)$$
\vskip .3cm
\noindent  Finally, make a special rotation $A_{3}$ around the first axis with
\vskip .3cm
$$\left(\matrix{
1\cr
\cr
{} &cos\theta_{3}  &-sin\theta_{3}\cr
{}\cr
{} &sin\theta_{3}  &~cos\theta_{3}\cr
}\right)\left(\matrix{
\left<{\eta_{1}\over{\eta_{1}}}\right>^{\prime\prime}\cr
\cr
0\cr
\cr
0\cr
}\right)=\left(\matrix{
\left<{\eta_{1}\over{\eta_{1}}}\right>^{\prime\prime}\cr
\cr
0\cr
\cr
0\cr
}\right),\left(\matrix{
1\cr
\cr
{} &cos\theta_{3}  &-sin\theta_{3}\cr
{}\cr
{} &sin\theta_{3}  &~cos\theta_{3}\cr
}\right)\left(\matrix{
\left<{\eta_{1}\over{\eta_{2}}}\right>^{\prime\prime}\cr
\cr
\left<{\eta_{2}\over{\eta_{2}}}\right>^{\prime\prime}\cr
\cr
\left<{\eta_{3}\over{\eta_{2}}}\right>^{\prime\prime}\cr
}\right)=\left(\matrix{
\left<{\eta_{1}\over{\eta_{2}}}\right>^{\prime\prime}\cr
\cr
\left<{\eta_{2}\over{\eta_{2}}}\right>^{\prime\prime\prime}\cr
\cr
0\cr
}\right),$$

$$\left(\matrix{
1\cr
\cr
{} &cos\theta_{3}  &-sin\theta_{3}\cr
{}\cr
{} &sin\theta_{3}  &~cos\theta_{3}\cr
}\right)\left(\matrix{
\left<{\eta_{1}\over{\eta_{3}}}\right>^{\prime\prime}\cr
\cr
\left<{\eta_{2}\over{\eta_{3}}}\right>^{\prime\prime}\cr
\cr
\left<{\eta_{3}\over{\eta_{3}}}\right>^{\prime\prime}\cr
}\right)=\left(\matrix{
\left<{\eta_{1}\over{\eta_{3}}}\right>^{\prime\prime\prime}\cr
\cr
\left<{\eta_{2}\over{\eta_{3}}}\right>^{\prime\prime\prime}\cr
\cr
\left<{\eta_{2}\over{\eta_{3}}}\right>^{\prime\prime\prime}\cr
}\right),\eqno(2.17)$$
\vskip .3cm
\noindent Hence, the special orthogonal matrix ${\bf Y}$ is
\vskip .3cm
$$Y=A_{3}A_{2}A_{1}.\eqno(2.18)$$
\vskip .3cm
\noindent It is obvious that the angle $\theta_{1}$ is fixed by
$\left<{\eta_{1}\over{\eta_{1}}}\right>$ and 
$\left<{\eta_{2}\over{\eta_{1}}}\right>$,  $\theta_{2}$ is fixed by
$\left<{\eta_{1}\over{\eta_{1}}}\right>^{\prime}$ and 
$\left<{\eta_{3}\over{\eta_{1}}}\right>$, and  $\theta_{3}$ is fixed by
$\left<{\eta_{2}\over{\eta_{2}}}\right>^{\prime\prime}$ and 
$\left<{\eta_{3}\over{\eta_{2}}}\right>^{\prime\prime}$. One can extend this special rotation  to
$m$-dimemnsional space,  and find that there are indeed unique solutions to $m$ rotational angles if the final
form (2.14) is selected. But we should point out that the choice of the rotation is not unique because
there are infinite number of solutions to the RWCs or WCs with outer multiplicity, which are all within the
$SO(m)$ transformations, and the number of elements in $SO(m)$ is infinite.
\vskip .3cm
   However, once the special rotation given in (2.14) is chosen, the resolution to the outer multiplicity
is thus fixed. There will no longer be arbitrariness for the RWCs with multiplicity except an over all phase
factor. We shall show that the over all phase factor can be chosen as 
\vskip .3cm
$$\left<{\eta_{i}\over{\eta_{i}}}\right>\geq 0\eqno(2.19)$$
\vskip .3cm
\noindent for $i=1,~2,~\cdots,~m$. Thus, the structure of the RWCs is determined completely.
\vskip .3cm
  One can easily prove that the WCs or RWCs after such transformation will not change the orthogonality
conditions. i. e., the orthogonality conditions given by (2.6) are still valid after transformation for
both ${\cal U}(4)\supset {\cal U}(3)$ and $SU(3)\supset U(2)$ cases. Firstly, all ${\cal U}(4)\supset {\cal U}(3)$
or $SU(3)\supset U(2)$ RWCs will undergo the same transformation ${\bf Y}$. One can verify that the
orthogonality conditions are still valid for them. $SU(3)$ WCs or 
$SU(3)\supset U(2)$ RWCs are a sub-set of those of ${\cal U}(4)$ or ${\cal U}(4)\supset {\cal U}(3)$ according
to the Schur-Weyl duality relation. Hence, the same conclusion applies to WCs of $SU(3)$ or RWCs of $SU(3)\supset
U(2)$ as well.
\vskip .3cm
   However, unlike Biedenharn's definition for WCs or RWCs , some symmetry properties of the
WCs or the RWCs will be changed. For example, the new RWCs do not satisfy the symmetry
property for
$1\leftrightarrow 2$ exchange of Biedenharn's due to the special orthogonal transformation ${\bf Y}$. We will
discuss this later in Sec. IV. 
\vskip .3cm
   Finally, we want to show what have been achieved after the special rotation ${\bf Y}$. If we arrange the RWCs of
${\cal U}(4)\supset {\cal U}(3)$ in terms of $m\times m$ matrix. The column is set by the outer multiplicity
label $\tilde{\eta}=\eta_{1},~\eta_{2},\cdots,~\eta_{m}$, while the row is set by the label $\eta$ in the irrep
$[m_{1},~m_{2}-\eta,~m_{3}-\mu_{2}+\eta ]$ for ${\cal U}(3)$, the RWCs will have the following structure
\vskip .3cm
$$\left(\left<{\tilde{\eta}\over{\eta}}\right>\right)=\left(
\matrix{
\times &0 &\cdots &\cdots&0\cr
\times &\times &0 &\cdots &0\cr
\times &\times &\times &0\cdots &0\cr
\cdots&\cdots\cr
\times &\times &\cdots &\times &0\cr
\times &\times &\cdots &\cdots&\times\cr}\right).\eqno(2.21)$$
\vskip .3cm
\noindent i.e., it just reflects the lower triangular structure of the RWCs with
multiplicity postulated by Braunschweig$^{[35]}$. Here, however, we have show that it is indeed possible to choose
such structure. Hecht in [22] argued that one can resolve the $SU(3)$ multiplicity problem simply by requiring a
similar lower triangular structure. Le Blanc and Rowe also pointed out that such resolution will becomes
{\it ipso facto} equivalent to a canonical labeling scheme$^{[27]}$. Actually,
One can also choose a upper triangular structure for these RWCs. Therefore, the structure of the
RWCs is also not unique, which depends on what kind of special transformation is chosen.
\vskip .3cm
   Now, let us recall Biedenharn's definition for canonical resolution to the outer multiplicity problem. In 
[6], `canonical' was used in the sense that there are no free choices involved in the solution of the
multiplicity. This explanation seems  incorrect because the choice involved in the resolution of the
multiplicity is not unique. This situation is quite the same as the definition of canonical basis for $U(n)$.
A canonical construction has to be explained as an equivalent class corresponding to the designation of a
particular $U(1)$, out of the set of all equivalent $U(1)$ groups, at each stage of the decomposition. While
the canonical resolution to the outer multiplicity has also to be explained as an equivalent class of solutions
with respect to the outer multiplicity labels  to be chosen, with which the WCs are mutually orthogonal.   
\vskip .3cm
  The special transformation given by (2.14) makes it possible to evaluate all RWCs of $SU(3)$ in the canonical
or noncanonical basis by using the recoupling approach. In the following, we will only discuss the RWCs for
$SU(3)$ in its canonical basis. The RWCs for $SU(3)$ in its noncanonical basis will be discussed elsewhere.
\vskip .5cm
\begin{centering}
{\large III.  A recoupling approach to the resolution of $SU(3)\supset U(2)$ RWCs }\\
\end{centering}
\vskip .4cm
  In this section, we want to demonstrate that the RWCs with multiplicity for $SU(3)\supset U(2)$, or
$SU(n)\supset U(n-1)$ in general, can be evaluated by using the accumulated results on the subject together
with the special choice of the transformation for a special set of RWCs, especially the analytical expression
for RWCs of $U(n)\supset U(n-1)$ with one irrep symmetric$^{[15]}$ given by Ali\v{s}auskas et al. The $U(3)\supset
U(2)$
RWCs of the same type was also obtained at the same period by Chacon et al$^{[5]}$ based on the canonical unit
tensor operator method proposed by Biedenharn et al. The result of Chacon's and that of Ali\v{s}auskas et al's
are the same including the phase factor. In order to make every step clear, we will divided this section
into several subsections.
\vskip .5cm
\noindent {\bf (a) A recoupling approach}
\vskip .3cm
   Using the analytical expressions of RWCs given by [15], one can construct the following expression for 
${\cal U}(4)$, and $SU(3)$, respectively, with the help of the building-up principle.$^{[36-37]}$
\vskip .3cm
$$\sum_{\xi}U_{\xi}\left(
\matrix{
(\lambda_{1}\mu_{1}) &[\lambda_{2}+\mu_{2}~0] &[\bar{m}]\cr
[\mu_{2}~0] &[m_{1}m_{2}m_{3}] &(\lambda_{2}\mu_{2})\cr}
\right)\left<\matrix{
(\lambda_{1}\mu_{1}) &(\lambda_{2}\mu_{2}) \cr
\rho_{1} &\rho_{2}\cr}\right\vert\left.
\matrix{
\xi~[m_{1}m_{2}m_{3}]\cr
\rho\cr}
\right>=$$

$$\sum\left<\matrix{
(\lambda_{1}\mu_{1}) &[\lambda_{2}+\mu_{2}~0] \cr
\rho_{1} &\rho_{2}^{\prime}\cr}\right\vert\left.
\matrix{
[\bar{m}]\cr
\bar{\rho}\cr}\right>\left<\matrix{
[\bar{m}] &[\mu_{2}~0] \cr
\bar{\rho} &\rho_{2}^{\prime\prime}\cr}\right\vert\left.
\matrix{
[m_{1}m_{2}m_{3}]\cr
\rho\cr}\right>\left<\matrix{
[\lambda_{2}+\mu_{2}~0] &[\mu_{2}~0] \cr
\rho_{2}^{\prime} &\rho_{2}^{\prime\prime}\cr}\right\vert\left.
\matrix{
(\lambda_{2}\mu_{2})\cr
\rho_{2}\cr}\right>,\eqno(3.1)$$
\vskip .3cm
\noindent where $U$ is unitary form of Racah coefficient for $SU(3)$ if $SU(3)$ case is considered, or is
for
${\cal U}(4)$ if we discuss the ${\cal U}(4)$ coupling case, and the sum on the rhs. is over 
$\rho_{2}^{\prime}$, $\rho_{2}^{\prime\prime}$ and $\bar{\rho}$. 
\vskip .3cm
   We will frequently use the following
abbreviated notations.
\vskip .3cm
$$U_{\xi}([\bar{m}])\equiv U_{\xi}\left(
\matrix{
(\lambda_{1}\mu_{1}) &[\lambda_{2}+\mu_{2}~0] &[\bar{m}]\cr
[\mu_{2}~0] &[m_{1}m_{2}m_{3}] &(\lambda_{2}\mu_{2})\cr}
\right)$$
\noindent  for the Racah coefficient,
\vskip .3cm
$$\left<{\tilde{\eta}\over{\eta}}\right>\equiv\left<
\matrix{
[\lambda_{1}+\mu_{1},~\mu_{1}] &[\lambda_{2}+\mu_{2},~\mu_{2}]\cr
[\lambda_{1}+\mu_{1},~\mu_{1}] &[\lambda_{2}+\mu_{2},~0]\cr
}
\right\vert\left.
\matrix{
\tilde{\eta}~[m_{1}m_{2}m_{3}]\cr
[m_{1},~m_{2}-\eta,~m_{3}-\mu_{2}+\eta ]\cr
}\right>$$
\vskip .3cm
\noindent for  a special set of ${\cal U}(4)\supset U(3)$ RWCs,
\vskip .3cm
$$\left<{\tilde{\eta}\over{\rho_{1}\rho_{2}\rho}}\right>\equiv\left<
\matrix{
[\lambda_{1}+\mu_{1},~\mu_{1}] &[\lambda_{2}+\mu_{2},~\mu_{2}]\cr
\rho_{1} &\rho_{2}\cr
}
\right\vert\left.
\matrix{
\tilde{\eta}~[m_{1}m_{2}m_{3}]\cr
\rho\cr
}\right>$$
\vskip .3cm
\noindent for $SU(3)$ WCs, or $SU(3)\supset U(2)$ RWCs if $\rho_{1}$, $\rho_{2}$, and $\rho$ are referred to as
the corresponding $U(2)$ labels,
\vskip .3cm
$$\left(\matrix{
\eta^{\prime}\cr
\eta\cr}\right)\equiv\left(\matrix{
[\lambda_{1}+\mu_{1}~\mu_{1}] &[\lambda_{2}+\mu_{2}~0]
&[m_{1}~m_{2}-\eta^{\prime}~m_{3}-\mu_{2}+\eta^{\prime}]\cr
[\lambda_{1}+\mu_{1}~\mu_{1}] &[\lambda_{2}+\mu_{2}~\mu_{2}] &[m_{1}~m_{2}~m_{3}]\cr
[\lambda_{1}+\mu_{1}] &[\lambda_{2}+\mu_{2}~0] &[m_{1}~m_{2}-\eta~m_{3}-\mu_{2}+\eta ]\cr}\right)$$
\vskip .3cm
\noindent for $\left({\cal U}_{4}\supset{\cal U}(3)\right)\star\left({\cal U}(4)\supset {\cal U}(3)\right)$
reduced coupling coefficient, and
\vskip .3cm
$$\left(\matrix{
\eta^{\prime}\cr
\rho_{1}\rho_{2}\rho\cr}\right)\equiv\left(\matrix{
[\lambda_{1}+\mu_{1}~\mu_{1}] &[\lambda_{2}+\mu_{2}~0]
&[m_{1}~m_{2}-\eta^{\prime}~m_{3}-\mu_{2}+\eta^{\prime}]\cr
[\lambda_{1}+\mu_{1}~\mu_{1}] &[\lambda_{2}+\mu_{2}~\mu_{2}] &[m_{1}~m_{2}~m_{3}]\cr
\rho_{1} &\rho_{2} &\rho\cr}\right)$$
\vskip .3cm
\noindent for $\left({\cal U}(4)\supset {\cal U}(3)\right)\star SU(3)$ coupling coefficient.
\vskip .3cm
   In ${\cal U}(4)\supset {\cal U}(3)$ case, we only need to consider a simpler case
\vskip .3cm
$$\sum_{\xi}U_{\xi}([\bar{m}])\left<
{\xi\over{\eta}}\right>=G([\bar{m}],\eta),\eqno(3.2)$$
\vskip .3cm
\noindent where
\vskip .3cm
$$G([\bar{m}],\eta)=\sum_{p_{1}p_{2}[\bar{m^{\prime}}]}
\left<\matrix{
[\lambda_{1}+\mu_{1}~\mu_{1}] &[\lambda_{2}+\mu_{2}~0]\cr
[\lambda_{1}+\mu_{1}~\mu_{1}] &p_{1}\cr
[\lambda_{1}+\mu_{1}~\mu_{1}] &0\cr}\right\vert\left.
\matrix{
[\bar{m}]\cr
[\bar{m^{\prime}}]\cr
[\lambda_{1}+\mu_{1}~\mu_{1}]\cr}\right>$$

$$\left<\matrix{
[\bar{m}] &[\mu_{2}~0]\cr
[\bar{m^{\prime}}] &p_{2}\cr
[\lambda_{1}+\mu_{1}~\mu_{1}] &0\cr}\right\vert\left.
\matrix{
[m_{1}m_{2}m_{3}]\cr
[m_{1}~m_{2}-\eta~m_{3}-\mu_{2}+\eta]\cr
[\lambda_{1}+\mu_{1}~\mu_{1}]\cr}\right>
\left<\matrix{
[\lambda_{2}~0] &[\mu_{2}~0]\cr
p_{1} &p_{2}\cr
}\right\vert\left.
\matrix{
[\lambda_{2}+\mu_{2}~\mu_{2}]\cr
[\lambda_{2}+\mu_{2}~0]\cr}\right>$$

$$\left<\matrix{
[\lambda_{1}+\mu_{1}~\mu_{1}] &[\lambda_{2}+\mu_{2}~0]\cr
[\lambda_{1}+\mu_{1}~\mu_{1}] &0\cr
}\right\vert\left.
\matrix{
[m_{1}~m_{2}-\eta~m_{3}-\mu_{2}+\eta]\cr
[\lambda_{1}+\mu_{1}~\mu_{1}]\cr}\right>^{-1}.\eqno(3.3)$$
\vskip .3cm
\noindent While in $SU(3)$ case, the following expression is of importance
\vskip .3cm
$$\sum_{\xi}U_{\xi}([\bar{m}])\left<\matrix{
(\lambda_{1}\mu_{1}) &(\lambda_{2}\mu_{2})\cr
\rho_{1} &\rho_{2}\cr
}\right\vert\left.
\matrix{
\xi~[m_{1}~m_{2}~m_{3}]\cr
\rho\cr}\right>=G([\bar{m}],\rho_{1}\rho_{2}\rho),\eqno(3.4)$$
\vskip .3cm
\noindent where
\vskip .3cm
$$\left<\matrix{
(\lambda_{1}\mu_{1}) &(\lambda_{2}\mu_{2})\cr
\rho_{1} &\rho_{2}\cr
}\right\vert\left.
\matrix{
\xi~[m_{1}~m_{2}~m_{3}]\cr
\rho\cr}\right>$$
\vskip .3cm
\noindent is the  WCs of $SU(3)$, and $G([\bar{m}],\rho_{1}\rho_{2}\rho)$
 is the expression given by the rhs. of (3.1).
\vskip .3cm
   Using these $G$ polynomials, one can easily construct the following coupling coefficients.
\vskip .3cm
$$\left(\matrix{
\eta^{\prime}\cr
\eta\cr}\right)=\sum_{[\bar{m}]}G([\bar{m}],\eta)G([\bar{m}],\eta^{\prime}),\eqno(3.5)$$
\vskip .3cm
\noindent which is a symmetric function with respect to the labels $\eta,~\eta^{\prime}$. What (3.5) gives
is nothing but reduced coupling coefficient of ${\cal U}(4)\star{\cal U}(4)$
\vskip .3cm
$$\left<
\left(\matrix{
[\lambda_{1}+\mu_{1}~\mu_{1}]\cr[\lambda_{1}+\mu_{1}~\mu_{1}]\cr[\lambda_{1}+\mu_{1}~\mu_{1}]\cr}\right);
\left(\matrix{
[\lambda_{2}+\mu_{2}~0]\cr[\lambda_{2}+\mu_{2}~\mu_{2}]\cr[\lambda_{2}+\mu_{2}~0]\cr}\right)\right\vert\left.
\left(\matrix{
[m_{1}~m_{2}-\eta^{\prime}~m_{3}-\mu_{2}+\eta^{\prime}]\cr
[m_{1}m_{2}m_{3}]\cr
[m_{1}~m_{2}-\eta~m_{3}-\mu_{2}+\eta]\cr}\right)\right>.\eqno(3.6)$$
\vskip .3cm
\noindent  Similarly, we have
\vskip .3cm
$$\left(\matrix{
\eta\cr\rho_{1}\rho_{2}\rho\cr}\right)=\sum_{[\bar{m}]}G([\bar{m}],\eta)G([\bar{m}],\rho_{1}\rho_{2}\rho),
\eqno(3.7)$$
\vskip .3cm
\noindent which is the following ${\cal U}(4)\star SU(3)$ coupling coefficient
\vskip .3cm
$$\left<
\left(\matrix{
[\lambda_{1}+\mu_{1}~\mu_{1}]\cr[\lambda_{1}+\mu_{1}~\mu_{1}]\cr\rho_{1}\cr}\right);
\left(\matrix{
[\lambda_{2}+\mu_{2}~0]\cr[\lambda_{2}+\mu_{2}~\mu_{2}]\cr\rho_{2}\cr}\right)\right\vert\left.
\left(\matrix{
[m_{1}~m_{2}-\eta~m_{3}-\mu_{2}+\eta]\cr
[m_{1}m_{2}m_{3}]\cr
\rho\cr}\right)\right>.\eqno(3.6)$$
\vskip .3cm
  As pointed out by Biedenharn et al$^{[2]}$, there are close relations between  RWCs of $U(n)$ and the coupling
coefficients of $U(n)\star U(n)$. They have also proved that such coupling coefficients can be expressed
in terms of the product of two $U(n)$ WCs with summation over the outer multiplicity labels. One can directly
verify by using the recoupling technique that
\vskip .3cm
$$\left(\matrix{
\eta^{\prime}\cr
\eta\cr}\right)=\sum_{[\bar{m}]}G([\bar{m}],\eta)G([\bar{m}],\eta^{\prime})
=\sum_{\xi}
\left<{\xi\over{\eta}}\right>\left<{\xi\over{\eta^{\prime}}}\right>\eqno(3.9)$$
\vskip .3cm
\noindent and
\vskip .3cm
$$\left(\matrix{
\eta\cr
\rho_{1}\rho_{2}\rho\cr}\right)=\sum_{[\bar{m}]}G([\bar{m}],\eta)G([\bar{m}],\rho_{1}\rho_{2}\rho)
=\sum_{\xi}
\left<{\xi\over{\eta}}\right>\left<{\xi\over{\rho_{1}\rho_{2}\rho_{3}}}\right>\eqno(3.10)$$
\vskip .3cm
\noindent {\bf (b) Explicit expressions of $G$ polynomials}
\vskip .3cm
   Using the recoupling technique, we can evaluate explicit expressions for the $G$ polynomials
defined in the above subsection with the known multiplicity-free RWCs of $U(n)\supset U(n-1)$ given
by Ali\v{s}auskas et al.
\vskip .3cm
$$G(k_{1}k_{2},\eta)=\left[
{(m_{3}-\mu_{2}+\eta)!^{2}(m_{2}-\eta+1)!^{2}(m_{1}+2)!\lambda_{1}!(\lambda-\mu_{1}-2k_{1}-k_{2}+2)
(\mu_{1}+k_{1}-k_{2}+1)\over{(\lambda_{2}+\mu_{2})!^{2}(\mu_{1}+\mu_{2}-m_{3}-\eta)!(\lambda_{2}+\mu_{2}+1)}}
\right.\times$$

$${(\lambda-k_{1}-k_{2}-\lambda_{1}-\mu_{1})!(\lambda-k_{1}-k_{2}-\mu_{1}+1)!
(m_{1}-m_{2}+1)!(m_{1}-m_{3}+2)!(m_{2}-m_{3}+1)\over{(\lambda_{1}-k_{1})!(\lambda_{1}+\mu_{1}-k_{2}+1)!
(\mu_{1}-k_{2})!(\mu_{2}-\eta)!(\lambda_{1}+\mu_{1}-m_{2}+\eta)!(\lambda_{1}+\mu_{1}+\mu_{2}-m_{3}-\eta+1)!
}}\times$$

$${(\lambda-k_{1}-k_{2}-m_{2})!(\lambda-k_{1}-k_{[2}-m_{3}+1)!(\lambda-k_{1}-k_{2}+2)!(\mu_{1}+k_{1}-m_{3})!
(\mu_{1}+k_{1}+1)!k_{1}!k_{2}!\over{(m_{1}-\lambda_{1}+k_{1}+k_{2})!(m_{1}-\mu_{1}-k_{1}+1)!(m_{1}-k_{2}+2)!
(m_{2}-\mu_{1}-k_{1})!(m_{2}+1)!m_{3}!(m_{3}-k_{2})!}}\times$$
\vskip .3cm
$$\left.{(\lambda_{2}+1)!\mu_{2}!(m_{2}-m_{3}-\eta)!(m_{1}-\lambda_{1}-\mu_{1})!(m_{1}-\mu_{1}+1)!(m_{2}-\mu_{1}-\eta)!\over{
(m_{1}-m_{2}+\eta+1)!(m_{1}-m_{3}+\mu_{2}-\eta+2)!\eta!(m_{2}-m_{3}+\mu_{2}-\eta+1)!}}\right]^{1/2}\times$$

$$\sum^{\mu_{2}}_{p=0}~\sum^{\max(m_{23})}_{m_{23}=\min(m_{23})}~\sum^{\max(m_{33})}_{m_{33}=\min(m_{33})}
(-)^{p}{p!(\lambda_{1}-m_{23})!(\lambda_{1}+\mu_{1}-m_{33}+1)!(\mu_{1}-m_{33})!(\lambda_{2}+\mu_{2}-p)!
\over{(\lambda-p-m_{23}-m_{33}-\lambda_{1}-\mu_{1})!(\lambda-p-m_{23}-m_{33}-\mu_{1}+1)!m_{23}!}}\times$$

$${(\lambda-p-2m_{23}-m_{33}-\mu_{1}+1)(\lambda-p-m_{23}-2m_{33}+2)(m_{23}-m_{33}+\mu_{1}+1)\over{
(\lambda-p-m_{23}-m_{33}+2)!(\mu_{1}+m_{23}+1)!m_{33}!}}\times$$

$$F_{3}\left(\matrix{
[\lambda-k_{1}-k_{2},\mu_{1}+k_{1},k_{2}]\cr
[\lambda-p-m_{23}-m_{33},m_{23}+\mu_{1},m_{33}]\cr}
;~\matrix{[m_{1}m_{2}m_{3}]\cr
[m_{1}~m_{2}-\eta~m_{3}-\mu_{2}+\eta]\cr}\right),\eqno(3.11)$$
\vskip .3cm
\noindent where 
\vskip .3cm
$$\lambda=\lambda_{1}+\mu_{1}+\lambda_{2}+\mu_{2},\eqno(3.12a)$$
\vskip .3cm
$$[\bar{m}]=[\lambda-k_{1}-k_{2},\mu_{1}+k_{1},k_{2}],\eqno(3.12b)$$
\vskip .3cm
\noindent the boundary conditions for  $m_{23}$,
and $m_{33}$ can be obtained by using the Littlewood rule for the Kronecker products involved and the
betweenness conditions for the decomposition $U(n)\downarrow U(n-1)$, from which we get
\vskip .3cm
$$\max(m_{23})=\min(\lambda_{1},k_{1},\lambda_{2}+\mu_{2}-p,m_{2}-\eta-\mu_{1}),$$

$$\min(m_{23})=\max(0,m_{2}-\eta-\mu_{1}-p,m_{3}-\mu_{1}-\mu_{2}+\eta),\eqno(13.12c)$$
\vskip .3cm
\noindent for fixed $p$, while 
\vskip .3cm
$$\max(m_{33})=\min(\lambda-p-m_{23}-m_{2}+\eta,m_{3}-\mu_{2}+\eta,\mu_{1},$$

$$\lambda_{2}+\mu_{2}-p-m_{23},
\lambda-p-m_{23}-\mu_{1}-k_{1},\lambda-p-m_{23}-\lambda_{1}-\mu_{1}),$$
\vskip .3cm
$$\min(m_{33})=\max(0,m_{2}+m_{3}+\mu_{1}-\mu_{2}-p+m_{23},k_{1}+k_{2}-p-m_{23},k_{2})\eqno(3.12d)$$
\vskip .3cm
\noindent for fixed $p$ and $m_{23}$, and 
\vskip .3cm
$$F_{3}\left(\matrix{
[h_{1}h_{2}h_{3}]\cr
[q_{1}q_{2}q_{3}]\cr};
~\matrix{[m_{1}m_{2}m_{3}]\cr
[n_{1}n_{2}n_{3}]\cr}\right)=\sum_{xyz}(-)^{x+y+z-q_{1}-q_{2}-q_{3}}\times$$

$${(x-y+1)(x-z+2)(y-z+1)(x-h_{3}+1)!\over{
(h_{2}-y)!(n_{2}-z+1)!(h_{3}-z)!(h_{1}-z+2)!(h_{1}-y+1)!}}\times$$

$${(q_{1}-y)!(q_{1}-z+1)!(q_{2}-z)!(x-n_{2})!(x-n_{3}+1)!(y-n_{3})!(y-h_{3})!\over{
(x-q_{1})!(x-q_{2}+1)!(x-q_{3}+2)!(y-q_{2})!(y-q_{3}+1)!(z-q_{3})!(n_{1}-x)!(n_{1}-y+1)!(n_{1}-z+2)!}}\times$$

$${(m_{1}-x)!(m_{1}-y+1)!(m_{1}-z+2)!(m_{2}-y)!(m_{2}-z+1)!(m_{3}-z)!(x-h_{2})!\over{
(n_{2}-y)!(n_{2}-z+1)!(n_{3}-z)!(x-m_{2})!(x-m_{3}+1)!(y-m_{3})!(h_{1}-x)!}}.\eqno(3.13)$$
\vskip .3cm
\noindent Similarly, we have
\vskip .3cm
$$G(k_{1}k_{2};[m_{12}m_{22}]~[m_{12}^{\prime}m_{22}^{\prime}][m_{12}^{\prime\prime}m_{22}^{\prime\prime}])
=\sum_{\xi}U_{\xi}(k_{1}k_{2})\left<\matrix{(\lambda_{1}\mu_{1}) &(\lambda_{2}\mu_{2})\cr
[m_{12}m_{22}] &[m_{12}^{\prime}m_{22}^{\prime}]\cr}\right\vert\left.\matrix{\xi~[m_{1}m_{2}m_{3}]\cr
[m_{12}^{\prime\prime}m_{22}^{\prime\prime}]}\right>=$$

$$\left[{(\lambda_{2}+1)!(\mu_{2}-m_{22}^{\prime})!(m_{1}-m_{2}+1)(m_{1}-m_{3}+2)(m_{2}-m_{3}+1)
(m_{12}^{\prime\prime}-m_{2})!
\over{(\lambda_{2}+\mu_{2}-m_{12}^{\prime})!(\lambda_{2}+\mu_{2}-m_{22}^{\prime}+1)!(m_{12}^{\prime}-\mu_{2})!
(m_{2}-k_{2}+1)!(m_{1}-m_{12}^{\prime\prime})!(m_{2}-m_{22}^{\prime\prime})!}}
\right.\times$$

$${(\lambda-k_{1}-k_{2}-m_{2})!(\mu_{1}+k_{1}-m_{3})!(\lambda-k_{1}-k_{2}-m_{3}+1)!(m_{22}^{\prime\prime}-m_{3})!
(m_{12}^{\prime\prime}-m_{3}+1)!\over{
(m_{1}-\lambda+k_{1}+k_{2})!(m_{2}-\mu_{1}-k_{1})!(m_{3}-k_{2})!(m_{1}-\mu_{1}-k_{1}+1)!(m_{1}-k_{2}+2)!}}
\times$$

$$\left.{(\mu_{1}-k_{2})!(\lambda_{1}+\mu_{1}-k_{2}+1)!(\lambda_{1}+\mu_{1}-m_{12})!(\mu_{1}-m_{22})!
(\lambda_{1}+\mu_{1}-m_{22}+1)!\over{
(\mu_{1}+k_{1}+1)!(m_{12}-\mu_{1})!m_{22}!(m_{12}+1)!(m_{1}-m_{22}^{\prime\prime}+1)!}}\right]^{1/2}
\sum^{\mu_{2}}_{p=0}~\sum^{\max(q)}_{q=\min(q)}\times$$

$$(-1)^{p-m_{22}^{\prime}}(\lambda_{2}+\mu_{2}+p-m_{12}^{\prime}
-m_{22}^{\prime})!((m_{12}^{\prime}-p)!(m_{12}^{\prime\prime}+m_{22}^{\prime\prime}-p-q-m_{12})!(q-m_{22})!)^{1/2}\times$$

$$\left({(m_{12}^{\prime\prime}
+m_{22}^{\prime\prime}-p-q-m_{22}+1)!
(m_{12}^{\prime\prime}+m_{22}^{\prime\prime}-p-2q+1)(p+q-m_{22}^{\prime\prime})!
(m_{22}^{\prime\prime}-q+1)!\over{(p-m_{22}^{\prime})!(m_{12}^{\prime\prime}-p-q)!(m_{12}^{\prime\prime}
+m_{22}^{\prime\prime}-p-q-\mu_{1}-k_{1})!(m_{12}-q)!}}\right) ^{1/2}\times$$

$$U\left(\matrix{[m_{12}m_{22}]
&[m_{12}^{\prime}+m_{22}^{\prime}-p~0] &[m_{12}^{\prime\prime}+m_{22}^{\prime\prime}-p-q,q]\cr
[p0] &[m_{12}^{\prime\prime}~m_{22}^{\prime\prime}] &[m_{12}^{\prime}~m_{22}^{\prime}]\cr}\right)\times$$

$$F_{2}\left(\matrix{[\lambda-k_{1}-k_{2},\mu_{1}+k_{1},k_{2}]
&[m_{1}m_{2}m_{3}]\cr
[m_{12}^{\prime\prime}+m_{22}^{\prime\prime}-p-q,q]
&[m_{12}^{\prime\prime}~m_{22}^{\prime\prime}]\cr}\right)\times$$

$$F_{2}\left(\matrix{[\lambda_{1}+\mu_{1}~\mu_{1}]
 &[\lambda-k_{1}-k_{2},\mu_{1}+k_{1},k_{2}]\cr
[m_{12}+m_{22}~0]
&[m_{12}^{\prime\prime}+m_{22}^{\prime\prime}-p-q,q]\cr}\right),\eqno(3.14)$$
\vskip .3cm
\noindent where 
\vskip .3cm
$$\max(q)=\min(\mu_{1}+k_{1},m_{12}^{\prime\prime}+m_{22}^{\prime\prime}-p-\mu_{1}-k_{1}),$$

$$\min(q)=\max(k_{2},m_{12}^{\prime\prime}+m_{22}^{\prime\prime}-p-\lambda+k_{1}+k_{2}),\eqno(3.15a)$$
\vskip .3cm
\noindent  $U$ is $SU(2)$ Racah coefficient in unitary form, and
\vskip .3cm
$$F_{2}\left(\matrix{[h_{1}h_{2}h_{3}]
&[m_{1}m_{2}m_{3}]\cr
[q_{1}q_{2}] 
&[n_{1}n_{2}]\cr}\right)=\sum_{xy}(-)^{x+y-q_{1}-q_{2}}~(x-y+1)\times$$

$${(x-n_{2})!(m_{1}-x)!(x-h_{2})!(x-h_{3}+1)!\over{
(x-q_{1})!(x-q_{2}+1)!(n_{1}-x)!(x-m_{2})!(x-m_{3}+1)!(h_{1}-x)!}}\times$$

$${(q_{1}-y)!(m_{1}-y+1)!(m_{2}-y)!(y-h_{3})!\over{
(y-q_{2})!(n_{2}-y)!(n_{1}-y+1)!(y-m_{3})!(h_{1}-y+1)!(h_{2}-y)!}}.
\eqno(3.15b)$$
\vskip .3cm
\noindent Using these expressions, (3.9) and (3.10) can be expressed explicitly as
\vskip .3cm
$$\left(\matrix{
\eta^{\prime}\cr
\eta\cr}\right)=\sum^{\max(k_{1})}_{k_{1}=\min(k_{1})}\sum^{\max(k_{2})}_{k_{2}=\min(k_{2})}
G(k_{1},k_{2},\eta)G(k_{1},k_{2},\eta^{\prime}),
\eqno(3.16a)$$
\vskip .3cm
$$\left(\matrix{
\eta\cr
\rho_{1}\rho_{2}\rho\cr}\right)=\sum^{\max(k_{1})}_{k_{1}=\min(k_{1})}\sum^{\max(k_{2})}_{k_{2}=\min(k_{2})}
G(k_{1},k_{2},\eta)G(k_{1}k_{2},\rho_{1}\rho_{2}\rho),\eqno(3.16b)
$$
\vskip .3cm
\noindent where
\vskip .3cm
$$\max(k_{1})=\min(m_{2}-\mu_{1},\lambda_{1},\lambda_{2}+\mu_{2}),$$

$$\min(k_{1})=\max(m_{3}-\mu_{1},m_{2}-\mu_{1}-\mu_{2},0),$$

$$\max(k_{2})=\min(\mu_{1},\lambda_{2}+\mu_{2}-k_{1},\lambda_{2}+\mu_{2}-k_{1}+\lambda_{1}+\mu_{1}-m_{2}),$$

$$\min(k_{2})=\max(m_{2}+m_{3}-\mu_{1}-\mu_{2}-k_{1}).\eqno(3.16c)$$
\vskip .5cm
\noindent {\bf (D) A recursive procedure for evaluation of all RWCs of $SU(3)\supset U(2)$}
\vskip .3cm
   We can now use the $G$ polynomials to construct ${\cal U}(4)\star{\cal U}(4)$ as well as
${\cal U}(4)\star SU(3)$ reduced coupling coefficients as given by (3.16a) and (3.16b), which can
be used to evaluate all RWCs of $SU(3)\supset U(2)$ after the special transformation given in Sec. II.
It can be proved that
\vskip .3cm
$$\sum_{\xi}\left<{\xi\over{\eta}}\right>\left<{\xi\over{\eta}}\right>\neq 0.\eqno(3.17)$$
\vskip .3cm
\noindent for any $\eta$. Firstly, if there exists an $\eta$ such that
\vskip .3cm
$$\sum_{\xi}\left<{\xi\over{\eta}}\right>\left<{\xi\over{\eta}}\right>= 0,\eqno(3.18)$$
\vskip .3cm
\noindent which requires that
\vskip .3cm
$$\left<{\xi\over{\eta}}\right>=0\eqno(3.19)$$
\vskip .3cm
\noindent for $\xi=\eta_{1},~\eta_{2},\cdots,~\eta_{m}$ because the condition (3.18) can be written as
$\vec{\eta}\cdot\vec{\eta}=0$ in the ${\bf R}^{m}$ space, which implies its components are all zero.
i. e.,
\vskip .3cm
$$\left<
\matrix{
[\lambda_{1}+\mu_{1},~\mu_{1}] &[\lambda_{2}+\mu_{2},~\mu_{2}]\cr
[\lambda_{1}+\mu_{1},~\mu_{1}] &[\lambda_{2}+\mu_{2},~0]\cr
}
\right\vert\left.
\matrix{
\xi~[m_{1}m_{2}m_{3}]\cr
[m_{1},~m_{2}-\eta,~m_{3}-\mu_{2}+\eta ]\cr}
\right>=0\eqno(3.20)$$
\vskip .3cm
\noindent for any multiplicity label $\xi$ and fixed $\eta$. However, using the building up principle,
one can deduce  that (3.20) is valid if and only if 
\vskip .3cm
$$G([\bar{m}],\eta)=0\eqno(3.21)$$
\vskip .3cm
\noindent where the expression of $G([\bar{m}],\eta)$ is given by (3.3).
In this case, we can always use unitarity condition for $U(n)\supset U(n-1)$ RWCs to prove that (3.21)
is satisfied only when one type of the WCs or RWCs involved are zero. However, one can verify from the explicit
expressions for these WCs or RWCs given by Ali\v{s}auskas et al$^{[15]}$ that (3.21) is satisfied only when the
irreps involved in the coupling do not satisfy Littlewood rule for the Kronecker products involved or betweenness
conditions for the decomposition. But these are all trivial cases and will not be considered. Hence, 
(3.17), generally, is valid for all non-trivial cases.
\vskip .3cm
   Hence, after the special transformation given in Sec. II, the rhs. of (3.9), generally, has $k$ non-zero
terms in the summation when a smaller $\eta$ which is a label in $[m_{1}~m_{2}-\eta~m_{3}-\mu_{2}+\eta]$ equals
to
$\eta_{k}$, namely
\vskip .3cm
$$\left(\matrix{
\eta_{k}\cr
\eta_{l}\cr}\right)=\sum_{\xi=\eta_{1}}^{\eta_{k}}
\left<{\xi\over{\eta_{k}}}\right>\left<{\xi\over{\eta_{l}}}\right>,~~~{\rm for}~~k\leq l,~k\leq m.\eqno(3.22)$$
\vskip .3cm
\noindent Using (3.22), we have
\vskip .3cm
$$\left<{\eta_{k}\over{\eta_{k}}}\right>=+\left[ \left(\matrix{
\eta_{k}\cr
\eta_{k}\cr}\right)-\sum_{\xi=\eta_{1}}^{\eta_{k-1}}
\left<{\xi\over{\eta_{k}}}\right>^{2}\right]^{1/2},\eqno(3.23)$$
\vskip .3cm
\noindent where the sign of (3.23) is always chosen positive for any $k$. The over all phase is thus fixed.
Then,
\vskip .3cm
$$\left<{\eta_{k}\over{\eta_{k}}}\right>\left<{\eta_{k}\over{\eta_{l}}}\right>= \left(\matrix{
\eta_{k}\cr
\eta_{l}\cr}\right)-\sum_{\xi=\eta_{1}}^{\eta_{k-1}}
\left<{\xi\over{\eta_{k}}}\right>\left<{\xi\over{\eta_{l}}}\right>\eqno(3.24)$$
\vskip .3cm
\noindent for $l>k$. It can be proved that (3.23) can not be zero.  Firstly,
$\left<\eta_{k}\over{\eta_{k}}\right>^{2}$ is square of the component of the vector $\vec{\eta_{k}}$ in ${\bf
R}^{m}$. Therefore, 
\vskip .3cm
$$\left<\eta_{k}\over{\eta_{k}}\right>^{2}\geq 0, ~~{\rm for}~~k\leq m.\eqno(3.25a)$$
\vskip .3cm
\noindent Secondly, if (3.23) is zero, (3.24) should  also be zero for any $l$ values. This only occurs when the
multiplicity equals to $k-1$. However, we assumed that the multiplicity $m\geq k$. Hence, (3.24) can not be zero
for $k\leq m$. Hence, 
\vskip .3cm
$$\left<\eta_{k}\over{\eta_{k}}\right>^{2}> 0, ~~{\rm for}~~k\leq m.\eqno(3.25b)$$
\vskip .3cm
   Thus, (3.23) and (3.24) allow us to calculate all the special RWCs of ${\cal U}(4)\supset {\cal U}(3)$
recursively,
\vskip .3cm
$$\left<\eta_{1}\over{\eta_{1}}\right>=\left(\matrix{
\eta_{1}\cr
\eta_{1}\cr}\right)^{1/2},$$

$$\left<\eta_{1}\over{\eta}\right>=\left(\matrix{
\eta_{1}\cr
\eta_{1}\cr}\right)^{-1/2}\left(\matrix{
\eta_{1}\cr
\eta\cr}\right),\eqno(3.26a)$$

$$\left<\eta_{2}\over{\eta}\right>=\left[\left(\matrix{
\eta_{2}\cr
\eta\cr}\right)-\left(\matrix{
\eta_{1}\cr
\eta_{1}\cr}\right)^{-1}\left(\matrix{
\eta_{1}\cr
\eta_{2}\cr}\right)\left(\matrix{
\eta_{1}\cr
\eta\cr}\right)\right]/\left<\eta_{2}\over{\eta_{2}}\right>,\eqno(3.26b)$$
\vskip .3cm
\noindent where
\vskip .3cm
$$\left<\eta_{2}\over{\eta_{2}}\right>=\left[\left(\matrix{
\eta_{2}\cr
\eta_{2}\cr}\right)-\left(\matrix{
\eta_{1}\cr
\eta_{1}\cr}\right)^{-1}\left(\matrix{
\eta_{1}\cr
\eta_{2}\cr}\right)^{2}\right]^{1/2}.\eqno(3.26c)$$
\vskip .3cm
$$\cdots~~\cdots$$
\noindent Once $\left<\eta_{k-1}\over{\eta}\right>$ for any $\eta$ are known from the $k-1$'th step,
$\left<\eta_{k}\over{\eta}\right>$ can be obtained by using (3.23) and (3.24). Thus, one obtains all
the special RWCs of ${\cal U}(4)\supset {\cal U}(3)$, which are important in determining the RWCs of $SU(3)
\supset U(2)$. 
\vskip .3cm
   Using (3.10) and all known special  RWCs of ${\cal U}(4)\supset {\cal U}(3)$, we can obtain
$SU(3)$ WCs or $SU(3)\supset U(2)$ RWCs
\vskip .3cm
$$\left<
\eta_{1}\over{
\rho_{1}\rho_{2}\rho}\right>=
\left<{\eta_{1}\over{\eta_{1}}}\right>^{-1}\left(\matrix{
\eta_{1}\cr
\rho_{1}\rho_{2}\rho\cr}\right),$$

$$\left<
\eta_{2}\over{
\rho_{1}\rho_{2}\rho}\right>=\left[\left(\matrix{
\eta_{2}\cr
\rho_{1}\rho_{2}\rho\cr}\right)-\left(\matrix{
\eta_{1}\cr
\eta_{1}\cr}\right)^{-1}\left(\matrix{
\eta_{1}\cr
\eta_{2}\cr}\right)\left(\matrix{
\eta_{1}\cr
\rho_{1}\rho_{2}\rho\cr}\right)\right]/\left<{\eta_{2}\over{\eta_{2}}}\right>,$$

$$\cdots~~\cdots$$

$$\left<
\eta_{k}\over{
\rho_{1}\rho_{2}\rho}\right>=\left[\left(\matrix{
\eta_{k}\cr
\rho_{1}\rho_{2}\rho\cr}\right)-\sum^{\eta_{k-1}}_{\eta^{\prime}=\eta_{1}}\left<
\eta^{\prime}\over{
\rho_{1}\rho_{2}\rho}\right>\left<
\eta^{\prime}\over{
\eta_{k}}\right>\right]/\left<{\eta_{k}\over{\eta_{k}}}\right>,~{\rm for~~}~k\leq m.\eqno(3.28)$$
\vskip .3cm
   It should be noted that (3.28) determines not only  $SU(3)$ WCs in the canonical basis, and
$SU(3)\supset U(2)$ RWCs, but also $SU(4)\supset U(3)$ RWCs for the same coupling.
\newpage
\noindent {\bf (E) Some algebraic expressions}\\
\vskip .3cm
   In this subsection, we will work out some algebraic expressions for $SU(3)\supset U(2)$ RWCs and
related Racah coefficients of $SU(3)$. Starting from $\eta=\eta_{1}$, and using (3.23), and (3.24),
we can obtain
\vskip .3cm
$$\left<
\eta_{1}\over{
\eta_{1}}\right>=\sum_{k_{1}k_{2}}G^{2}(k_{1},k_{2},\eta_{1}),$$

$$\left<
\eta_{1}\over{
\eta_{1+i}}\right>=\left<
\eta_{1}\over{
\eta_{1}}\right>^{-1}\sum_{k_{1}k_{2}}G(k_{1}k_{2},\eta_{i+1})G(k_{1}k_{2},\eta_{1}),$$

$$\left<
\eta_{2}\over{
\eta_{2}}\right>^{2}=\left<
\eta_{1}\over{
\eta_{1}}\right>^{-2}\sum^{1}_{i=0}\sum_{k_{1}k_{2}k^{\prime}_{1}k^{\prime}_{2}}(-)^{[i/2]}
G(k_{1}k_{2},\eta_{2})
G(k_{1}k_{2},\eta_{2-i})G(k^{\prime}_{1}k^{\prime}_{2},\eta_{1+i})G(k^{\prime}_{1}k^{\prime}_{2},\eta_{1}),$$

$$\cdots~~\cdots$$

$$\left<
\eta_{3}\over{
\eta_{3}}\right>^{2}=\left<
\eta_{1}\over{
\eta_{1}}\right>^{-5}\left<
\eta_{2}\over{\eta_{2}}\right>^{-1}\sum^{2}_{i=0}{}^{\prime}\sum^{1}_{j=0}{}^{\prime}\sum^{1}_{k=0}{}^{\prime}
\sum_{k_{l},p_{l},q_{l},n_{l}}(-)^{[i/2]+[j/2]+[k/2]}
G(k_{1}k_{2},\eta_{3})G(k_{1}k_{2},\eta_{3-i-j})\times$$

$$G(p_{1}p_{2},\eta_{1+i})G(p_{1}p_{2},\eta_{1})G(q_{1}q_{2},\eta_{2+j})G(q_{1}q_{2},\eta_{2-k})
G(n_{1}n_{2},\eta_{1+k})G(n_{1}n_{2},\eta_{1}),\eqno(3.29)$$
\vskip .3cm
\noindent where the primes on the summation signs indicate that the sums should be restricted by
\vskip .3cm
$$i+j+k\leq 2,\eqno(3.30)$$
\vskip .3cm
\noindent and
\vskip .3cm
$$[x]=\left\{
\begin{array}{l}
x~~{\rm if}~x~{\rm is~an ~integer,}\\
2x~~{\rm if}~x~{\rm is~ an~ half-integer}.
\end{array}\right.\eqno(3.31)$$
\vskip .3cm
\noindent The expression will become more complicated with increasing of $k$ in $\eta_{k}$. 
Once  $\left<\eta_{k}\over{\eta_{k}}\right>$ for $k=1,~2,~\cdots,~m$ are known, one can similarly
get the WCs or RWCs of $SU(3)\supset U(2)$ with multiplicity. 
\vskip .3cm
$$\left<\eta_{k}\over{\rho_{1}\rho_{2}\rho}\right>=
\left(\prod^{k-1}_{t=0}\left<\eta_{t}\over{\eta_{t}}\right>^{-1}{\sum_{i_{t}=0}^{k-t}}{}^{\prime}\right)
\sum_{k_{1}k_{2}}G(k_{1},k_{2},\rho_{1}\rho_{2}\rho)G(k_{1},k_{2},\eta_{k-\sum^{k-1}_{j=1}i_{j}})\times$$

$$(-)^{\sum^{k-1}_{p=1}[i_{p}/2]}\prod^{k-1}_{q=1}\left<\eta_{q}\over{\eta_{q+i_{p}}}\right>,\eqno(3.32)$$
\vskip .3cm
\noindent where the prime on the summation sum indicates that the condition
\vskip .3cm
$$\sum_{t=0}^{k-1}i_{t}\leq k-1\eqno(3.33)$$
\vskip .3cm
\noindent  should be satisfied. Similarly, we define
\vskip .3cm
$$V({\eta_{i}\over{\eta_{j}}})=\left<\eta_{i}\over{\eta_{i}}\right>^{-1}
\left<\eta_{i}\over{\eta_{j}}\right>.\eqno(3.34a)$$
\vskip .3cm
\noindent The Racah coefficient of $SU(3)$ can be expressed as
\vskip .3cm
$$R_{\eta_{1}}([\bar{m}])\left<\eta_{1}\over{\eta_{1}}\right>=G([\bar{m}],\eta_{1}),$$

$$R_{\eta_{2}}([\bar{m}])\left<\eta_{2}\over{\eta_{2}}\right>=G([\bar{m}],\eta_{2})-
V({\eta_{1}\over{\eta_{2}}})G([\bar{m}],\eta_{1}),$$

$$\cdots~~\cdots$$

$$R_{\eta_{j}}([\bar{m}])\left<\eta_{j}\over{\eta_{j}}\right>=G([\bar{m}],\eta_{j})+
\sum^{j-1}_{i_{1}=1}\sum^{j-1}_{k=1}(-)^{k}\sum_{i_{1}\leq i_{2}\leq\cdots\leq i_{k}<j}
G([\bar{m}],\eta_{i_{1}})\times$$

$$V({\eta_{i_{k}}\over{\eta_{j}}})\prod^{k-1}_{l=1}V({\eta_{i_{l}}\over{\eta_{i_{l+1}}}}).\eqno(3.34b)$$
\vskip .5cm
\begin{centering} 
{\large IV. Some features of the  new RWCs for $SU(3)\supset U(2)$}\\
\end{centering}
\vskip .3cm
   In this section, we will discuss symmetry properties of WCs of $SU(3)$ in the canonical basis.  In Sec. III,
we have chosen the absolute phase with
\vskip .3cm
$$\left<
\matrix{
[\lambda_{1}+\mu_{1},~\mu_{1}] &[\lambda_{2}+\mu_{2},~\mu_{2}]\cr
[\lambda_{1}+\mu_{1},~\mu_{1}] &[\lambda_{2}+\mu_{2},~0]\cr}
\right\vert\left.
\matrix{
\eta~[m_{1}m_{2}m_{3}]\cr
[m_{1},~m_{2}-\eta,~m_{3}-\mu_{2}+\eta ]\cr}
\right>\geq 0.\eqno(4.1)$$
\vskip .3cm
\noindent Then, the relative phase is completely determined by the recursion relations (3.23) and (3.24).
When $\mu_{2}=0$, one can check that our phase choice is consistent with that of [15], which
is the same as that of [5] with the same phase structure defined by Biedenharn et al[10] as it should be because
we use multiplicity-free RWCs of [5,15]. Hence, in order to discuss symmetry properties of the SU(3) WCs, we
can expand the WCs obtained in this paper in terms of those defined in [10]
\vskip .3cm

$$\left<
\matrix{
(\lambda_{1}\mu_{1}) &(\lambda_{2}\mu_{2})\cr
\rho_{1} &\rho_{2}\cr}
\right\vert\left.\matrix{
\eta~[m_{1}m_{2}m_{3}]\cr
\rho\cr}\right>=\sum_{\xi}y(\xi,\eta)\left[
\matrix{
(\lambda_{1}\mu_{1}) &(\lambda_{2}\mu_{2})\cr
\rho_{1} &\rho_{2}\cr}
\right\vert\left.\matrix{
\xi~[m_{1}m_{2}m_{3}]\cr
\rho\cr}\right],\eqno(4.2)$$
\vskip .3cm
\noindent where the coupling coefficients on the rhs. in square brackets are the WCs defined in [10],
and $y(\xi,\eta)$ is the corresponding special transformation matrix elements. Using the orthogonality relations
for $y(\xi,\eta)$'s, and symmetry properties of the WCs discussed in [10], we can prove that 
\vskip .3cm
$$\left<
\matrix{
(\lambda_{1}\mu_{1}) &(\lambda_{2}\mu_{2})\cr
\rho_{1} &\rho_{2}\cr}
\right\vert\left.\matrix{
\eta^{\prime}~[m_{1}m_{2}m_{3}]\cr
\rho\cr}\right>=\sum_{\eta}\sum_{\xi}y(\xi,\eta)y(\xi,\eta^{\prime})(-)^{\phi(\xi)}\left<
\matrix{
(\lambda_{2}\mu_{2}) &(\lambda_{1}\mu_{1})\cr
\rho_{2} &\rho_{1}\cr}
\right\vert\left.\matrix{
\eta~[m_{1}m_{2}m_{3}]\cr
\rho\cr}\right>,\eqno(4.2)$$
\vskip .3cm
\noindent where the phase factor $(-)^{\phi(\xi)}$ comes from
\vskip .3cm
$$\left[
\matrix{
(\lambda_{1}\mu_{1}) &(\lambda_{2}\mu_{2})\cr
\rho_{1} &\rho_{2}\cr}
\right\vert\left.\matrix{
\xi~[m_{1}m_{2}m_{3}]\cr
\rho\cr}\right]=(-)^{\phi(\xi)}\left[
\matrix{
(\lambda_{2}\mu_{2}) &(\lambda_{1}\mu_{1})\cr
\rho_{2} &\rho_{1}\cr}
\right\vert\left.\matrix{
\xi~[m_{1}m_{2}m_{3}]\cr
\rho\cr}\right]\eqno(4.3)$$
\vskip .3cm
\noindent given in [10], 
\vskip .3cm
$$\phi(\xi)=\Gamma_{12}-\Gamma_{11}+\mu_{1}+\mu_{2}-m_{2}+m_{1}.\eqno(4.4)$$
\vskip .3cm
\noindent The multiplicity label $\xi$ in this case can be regarded as 
\vskip .3cm
$$\xi=\Gamma_{12}-\Gamma_{11},\eqno(4.5)$$
\vskip .3cm
\noindent where $\Gamma_{ij}$ are multiplicity labels from the upper pattern of $SU(3)$ defined
by Biedenharn et al. 
\vskip .3cm
   It is obvious that 
$$Z_{\eta\eta^{\prime}}=Z\left(\matrix{
(\lambda_{1}\mu_{1}) &(0) &(\lambda_{1}\mu_{1})~\eta^{\prime}\cr
(\lambda_{2}\mu_{2}) &[m_{1}m_{2}m_{3}] &(\lambda_{2}\mu_{2})~\eta\cr}\right)
=\sum_{\xi}y(\xi,\eta)y(\xi,\eta^{\prime})(-)^{\phi(\xi)},\eqno(4.6)$$
\vskip .3cm
\noindent where $Z_{\eta\eta^{\prime}}$ is a special $Z$ coefficient$^{[38]}$ defined by Millener, which
transforms the coupling coefficients from the coupling $(\lambda_{1}\mu_{1})\times (\lambda_{2}\mu_{2})$
to $(\lambda_{2}\mu_{2})\times (\lambda_{1}\mu_{1})$. (4.2) can only be simplified when the coupling is
multiplicity-free. In this case, the transformation matrix $Y$ is $1\times 1$ with $y(\xi,\eta)=1$ for
fixed $\xi$ and $\eta$, and
\vskip .3cm
$$Z_{\eta\eta^{\prime}}=\delta_{\eta\eta^{\prime}}(-)^{\phi(\xi)},\eqno(4.7)$$
\vskip .3cm
\noindent where $\xi$ can be expressed in terms of $\lambda_{1},~\mu_{1},~\lambda_{2},~\mu_{2}$, and
$m_{i}$ with $i=1,~2,~3$.  
\vskip .3cm
 Similarly, we obtain the following symmetry properties for the WCs of $SU(3)$
\vskip .3cm
$$\left<
\matrix{
(\mu_{1}\lambda_{1}) &(\mu_{2}\lambda_{2})\cr
\tilde{\rho}_{1} &\tilde{\rho}_{2}\cr}
\right\vert\left.\matrix{
\eta^{\prime}~[-m_{3}~-m_{2}~-m_{1}]\cr
\tilde{\rho}\cr}\right>=$$

$$\sum_{\eta}Z_{\eta\eta^{\prime}}(-)^{m_{1}-m_{3}}\left<
\matrix{
(\lambda_{1}\mu_{1}) &(\lambda_{2}\mu_{2})\cr
\rho_{1} &\rho_{2}\cr}
\right\vert\left.\matrix{
\eta~[m_{1}m_{2}m_{3}]\cr
\rho\cr}\right>,\eqno(4.8a)$$
\vskip .3cm
\noindent where $\tilde{\rho}$ is the conjugation of $\rho$ defined by 
\vskip .3cm
$$\vert\tilde{\rho}>\equiv\vert\tilde{m}>=\left\vert\matrix{
-m_{22}~-m_{12}\cr
-m_{11}\cr}\right>,\eqno(4.8b)$$
\vskip .3cm
\noindent and 
\vskip .3cm
$$\left<
\matrix{
(\lambda_{1}\mu_{1}) &(\lambda_{2}\mu_{2})\cr
\rho_{1} &\rho_{2}\cr}
\right\vert\left.\matrix{
\eta^{\prime}~[m_{1}m_{2}m_{3}]\cr
\rho\cr}\right>=
\left[{\dim([m_{1}m_{2}m_{3}])\over{\dim((\lambda_{1}\mu_{1}))}}\right]^{1/2}(-)^{\phi_{3}-\phi_{1}}\times$$

$$\sum_{\eta}Z_{\eta\eta^{\prime}}\left<
\matrix{
[m_{1}m_{2}m_{3}] &(\mu_{2}\lambda_{2})\cr
\rho &\tilde{\rho}_{2}\cr}
\right\vert\left.\matrix{
\eta~(\lambda_{1}\mu_{1})\cr
\rho_{1}\cr}\right>,\eqno(4.9)$$
\vskip .3cm
\noindent where
\vskip .3cm
$$\phi_{3}=m^{\prime\prime}_{11}-m^{\prime\prime}_{12}-m_{22}^{\prime\prime}+m_{1},$$

$$\phi_{1}=m_{11}-m_{12}-m_{22}+\lambda_{1}+\mu_{1},\eqno(4.10)$$
\vskip .3cm
\noindent and
$$\rho=\left(\begin{array}{l}
[m_{12}^{\prime\prime}~m_{22}^{\prime\prime}]\\
~~~~m_{11}^{\prime\prime}
\end{array}\right),~~\rho_{1}=\left(\begin{array}{l}
[m_{12}~m_{22}]\\
~~~~m_{11}
\end{array}\right).\eqno(4.11)$$
\vskip .3cm
  Now, we give some examples for the coupling $[21]\times [21]\downarrow [321]$, which show us
main features of the new RWCs. We use the following notations.
\vskip .3cm
$$<{0\over{0}}>
=\left<\matrix{[21] &[21]\cr
[21] &[20]\cr}\right\vert\left.\matrix{
\eta=0~[321]\cr
~~~~~~~~[32]\cr}\right>,~<{1\over{0}}>
=\left<\matrix{[21] &[21]\cr
[21] &[20]\cr}\right\vert\left.\matrix{
\eta=1~[321]\cr
~~~~~~~~[32]\cr}\right>$$

$$<{0\over{1}}>
=\left<\matrix{[21] &[21]\cr
[21] &[20]\cr}\right\vert\left.\matrix{
\eta=0~[321]\cr
~~~~~~~~[311]\cr}\right>,~<{1\over{1}}>
=\left<\matrix{[21] &[21]\cr
[21] &[20]\cr}\right\vert\left.\matrix{
\eta=1~[321]\cr
~~~~~~~~[311]\cr}\right>.\eqno(4.12)$$
\vskip .3cm
\noindent  Using the recursion relations given by (3.23), and (3.24), we can easily get
\vskip .3cm
$$\left(\matrix{
<{0\over{0}}>  &<{1\over{0}}>\cr
<{0\over{1}}>  &<{1\over{1}}>\cr}\right)=
\left(\matrix{
\sqrt{7\over{10}}  &0\cr
-\sqrt{1\over{42}} &\sqrt{10\over{21}}\cr}\right).\eqno(4.13)$$
\vskip .3cm
\noindent Then, Tables I and II can be worked out by using (3.28). In these tables, the upper parts
are taken from de Swart [39], while the lower parts are derived by the new method. These two types of RWCs
with multiplicity two can be transformed with each other by a two dimensional rotation. Furthermore, one can check
that the new RWCs still satisfy the orthogonality conditions for $SU(3)$ RWCs given by (2.6).
\vskip .5cm
\begin{centering}
{\large V. Reduced auxiliary Wigner coefficients for $SU(3)\supset U(2)$}\\
\end{centering}
\vskip .4cm
   In [11], they used also $U(4)$ complementary group to label the multiplicity labels of $SU(3)$. However,
the $U(4)$ group in that case is labeled in a noncanonical $U(2)\times U(2)$ chain.  They also pointed out that
the so-called Auxiliary Wigner Coefficient (AWC) of $SU(3)$ can be calculated from a $U(4)\star SU(3)$ scalar
product. The reduced AWCs satisfy
\vskip .3cm
$$\sum_{\lambda_{2}\mu_{2}~q_{i}^{\prime}q_{i}^{\prime\prime}}\left<\matrix{(\lambda_{1}\mu_{1})
&(\lambda_{2}\mu_{2})\cr [q_{1}^{\prime}q_{2}^{\prime}] &[q_{1}^{\prime\prime}q_{2}^{\prime\prime}]\cr}\right\vert
\left.\matrix{[m_{1}m_{2}m_{3}]\cr [q_{1}q_{2}]\cr};[u_{1}u_{2}u_{3}]\right>\left<\matrix{(\lambda_{1}\mu_{1})
&(\lambda_{2}\mu_{2})\cr [q_{1}^{\prime}q_{2}^{\prime}] &[q_{1}^{\prime\prime}q_{2}^{\prime\prime}]\cr}\right\vert
\left.\matrix{[\bar{m}_{1}\bar{m}_{2}\bar{m}_{3}]\cr [q_{1}q_{2}]\cr};[\bar{u}_{1}\bar{u}_{2}\bar{u}_{3}]\right>$$

$$=\prod_{i}\delta_{\bar{u}_{i}u_{i}}\delta_{\bar{m}_{i}m_{i}},\eqno(5.1)$$

$$\sum_{u_{i}m_{i}}\left<\matrix{(\lambda_{1}\mu_{1})
&(\lambda_{2}\mu_{2})\cr [q_{1}^{\prime}q_{2}^{\prime}] &[q_{1}^{\prime\prime}q_{2}^{\prime\prime}]\cr}\right\vert
\left.\matrix{[m_{1}m_{2}m_{3}]\cr [q_{1}q_{2}]\cr};[u_{1}u_{2}u_{3}]\right>
\left<\matrix{(\lambda_{1}\mu_{1})
&(\lambda_{2}\mu_{2})\cr [\bar{q}_{1}^{\prime}\bar{q}_{2}^{\prime}]
&[\bar{q}_{1}^{\prime\prime}\bar{q}_{2}^{\prime\prime}]\cr}\right\vert
\left.\matrix{[m_{1}m_{2}m_{3}]\cr [q_{1}q_{2}]\cr};[u_{1}u_{2}u_{3}]\right>$$

$$=\prod_{i}\delta_{\bar{q}^{\prime\prime}_{i}q^{\prime\prime}_{i}}\delta_{\bar{q}_{i}q^{\prime}_{i}},\eqno(5.2)$$
\vskip .3cm
\noindent where $[u_{1}u_{2}u_{3}]$ are labels of ${\cal U}(3)$ which is the subgroup of $U(4)$. We can show that
such AWCs are equivalent to what will be constructed by using the recoupling approach. Firstly, we assume the irrep
$(\lambda_{2}\mu_{2})$ is a coupled irrep, namely
\vskip .3cm
$$\left\vert\matrix{
(\lambda_{2}\mu_{2})\cr
\rho_{2}\cr}\right>=\sum_{\rho^{\prime}_{2}\rho_{2}^{\prime\prime}}
\left<\matrix{(\lambda_{2}^{\prime}+\mu_{2}^{\prime}0) &(\mu_{2}^{\prime}0)\cr
\rho^{\prime}_{2} &\rho_{2}^{\prime\prime}\cr}\right\vert\left.
\matrix{(\lambda_{2}\mu_{2})\cr
\rho_{2}\cr}\right>\left\vert\matrix{
\lambda_{2}^{\prime}+\mu_{2}^{\prime}\cr
\rho_{2}^{\prime}\cr};\matrix{\mu_{2}^{\prime}\cr
\rho_{2}^{\prime\prime}\cr}\right>,\eqno(5.3)$$
\vskip .3cm
\noindent where $\rho_{2},~\rho_{2}^{\prime},~\rho_{2}^{\prime\prime}$ are sublabels of $SU(3)$ in the canonical 
basis. Then, a special type of final coupled basis vectors can be written as
\vskip .3cm
$$\left\vert ((\lambda_{1}\mu_{1}),(\lambda_{2}^{\prime}+\mu_{2}^{\prime}0))
[u_{1}u_{2}u_{3}],(\mu_{2}^{\prime}0),
\matrix{[m_{1}m_{2}m_{3}]\cr
\rho\cr}\right>\equiv\left\vert\matrix{
[m_{1}m_{2}m_{3}]\cr
\rho\cr};[u_{1}u_{2}u_{3}]\right>=$$

$$\sum\left<\matrix{
(\lambda_{1}\mu_{1}) &(\lambda^{\prime}_{2}+\mu_{2}^{\prime}0)\cr
\rho_{1} &\rho_{2}^{\prime}\cr}\right\vert\left.
\matrix{[u_{1}u_{2}u_{3}]\cr
\bar{\rho}\cr}\right>\left<\matrix{
[u_{1}u_{2}u_{3}] &(\mu^{\prime}_{2}0)\cr
\bar{\rho} &\rho_{2}^{\prime\prime}\cr}\right\vert\left.
\matrix{[m_{1}m_{2}m_{3}]\cr
\rho\cr}\right>\times$$

$$\left<\matrix{
(\lambda^{\prime}_{2}+\mu_{2}^{\prime}0) &(\mu^{\prime}_{2}0)\cr
\rho^{\prime}_{2} &\rho_{2}^{\prime\prime}\cr}\right\vert\left.
\matrix{(\lambda_{2}\mu_{2})\cr
\rho_{2}\cr}\right>\left\vert\matrix{
(\lambda_{1}\mu_{1})\cr
\rho_{1}\cr};\matrix{
(\lambda_{2}^{\prime}+\mu_{2}^{\prime}0)\cr
\rho_{2}^{\prime}\cr};\matrix{
(\mu_{2}^{\prime}0)\cr
\rho_{2}^{\prime\prime}\cr}\right>,\eqno(5.4)$$
\vskip .3cm
\noindent where the sum is over $\rho_{1},~\rho_{2}^{\prime},~\rho_{2}^{\prime\prime},~\bar{\rho}$, and the WCs
involved in the summation are all given by Ali\v{s}auskas et al [15] and Chacon et al [5]. 
\vskip .3cm
   The AWCs can be evaluated by the norm
\vskip .3cm
$$\left<\matrix{
(\lambda_{1}\mu_{1})\cr
\rho_{1}\cr};\matrix{(\lambda_{2}\mu_{2})\cr
\rho_{2}\cr}\right\vert\left.
\matrix{[m_{1}m_{2}m_{3}]\cr \rho\cr};[u_{1}u_{2}u_{3}]\right>=$$

$$\sum_{\rho_{2}^{\prime}\rho_{2}^{\prime\prime}\bar{\rho}}\left<\matrix{
(\lambda_{1}\mu_{1}) &(\lambda^{\prime}_{2}+\mu_{2}^{\prime}0)\cr
\rho_{1} &\rho_{2}^{\prime}\cr}\right\vert\left.
\matrix{[u_{1}u_{2}u_{3}]\cr
\bar{\rho}\cr}\right>\left<\matrix{
[u_{1}u_{2}u_{3}] &(\mu^{\prime}_{2}0)\cr
\bar{\rho} &\rho_{2}^{\prime\prime}\cr}\right\vert\left.
\matrix{[m_{1}m_{2}m_{3}]\cr
\rho\cr}\right>\times$$

$$\left<\matrix{
(\lambda^{\prime}_{2}+\mu_{2}^{\prime}0) &(\mu^{\prime}_{2}0)\cr
\rho^{\prime}_{2} &\rho_{2}^{\prime\prime}\cr}\right\vert\left.
\matrix{(\lambda_{2}\mu_{2})\cr
\rho_{2}\cr}\right>.\eqno(5.5)$$
\vskip .3cm
\noindent One can check that the labels $u_{1},u_{2},u_{3}$ take the same values as those given by  Brody et al
[11]. One can also  get reduced AWCs as follows.
\vskip .3cm
$$\left<\matrix{
(\lambda_{1}\mu_{1}) &(\lambda_{2}\mu_{2})\cr
[q_{1}q_{2}] &[q^{\prime}_{1}q_{2}^{\prime}]\cr}\right\vert\left.
\matrix{[m_{1}m_{2}m_{3}]\cr
[q_{1}^{\prime\prime}q^{\prime\prime}_{2}]\cr};[u_{1}u_{2}u_{3}]\right>=
\sum_{p~\bar{q}_{1}\bar{q}_{2}}
\left<\matrix{
(\lambda_{1}\mu_{1}) &(\lambda^{\prime}_{2}+\mu_{2}^{\prime}0)\cr
[q_{1}q_{2}] &[\lambda_{2}^{\prime}+\mu_{2}^{\prime}-p0]\cr}\right\vert\left.
\matrix{[u_{1}u_{2}u_{3}]\cr
[\bar{q}_{1}\bar{q}_{2}]\cr}\right>\times$$

$$\left<\matrix{
[u_{1}u_{2}u_{3}] &(\mu^{\prime}_{2}0)\cr
[\bar{q}_{1}\bar{q}_{2}] &[p0]\cr}\right\vert\left.
\matrix{[m_{1}m_{2}m_{3}]\cr
[q_{1}^{\prime\prime}q^{\prime\prime}_{2}]\cr}\right> \left<\matrix{
(\lambda^{\prime}_{2}+\mu_{2}^{\prime}0) &(\mu^{\prime}_{2}0)\cr
[\lambda_{2}^{\prime}+\mu_{2}^{\prime}-p0] &[p0]\cr}\right\vert\left.
\matrix{(\lambda_{2}\mu_{2})\cr
[q_{1}^{\prime}q_{2}^{\prime}]\cr}\right>\times$$

$$U\left(
\matrix{[q_{1}q_{2}] &[\lambda_{2}^{\prime}+\mu_{2}^{\prime}-p0] &[\bar{q}_{1}\bar{q}_{2}]\cr
[p0] &[q_{1}^{\prime\prime}q^{\prime\prime}_{2}] &[q_{1}^{\prime}q^{\prime}_{2}]\cr}\right).\eqno(5.6)$$
\vskip .3cm
\noindent One can verify that the reduced AWCs given by (5.5) indeed satisfy the orthogonality relations
(5.1) and (5.2). Using the explicit expression of the multiplicity -free RWCs involved in the sum, one can
get a closed algebraic expression for the reduced AWCs. Such AWCs may also be useful, especially in 
two-particle coupling problems. It should be pointed out that though the result (5.6) is the same as required
by Brody et al, the labeling scheme is, after all, not the same. The multiplicity labels of their AWCs are
specified by a $U(4)$ subgroup $U(3)$, while they are now specified by  the irrep of $SU(3)$ 
coupled from the first two irreps.
\vskip .3cm
   Using analytical expressions for the multiplicity-free RWCs  given in [15], we obtain the following
algebraic expression for the reduced AWCs  of $SU(3)\supset U(2)$
\vskip .3cm

$$\left<\matrix{
(\lambda_{1}\mu_{1}) &(\lambda^{\prime}_{2}+\mu^{\prime}_{2};\mu_{2}^{\prime})(\lambda_{2}\mu_{2})\cr
[m_{1}m_{2}] &[m^{\prime}_{1}m_{2}^{\prime}]\cr}\right\vert\left.
\matrix{[m_{1}m_{2}m_{3}]\cr
[m_{1}^{\prime\prime}m^{\prime\prime}_{2}]\cr};[u_{1}u_{2}u_{3}]\right>=\delta_{\lambda^{\prime}_{2}+
2\mu_{2}^{\prime},\lambda_{2}+2\mu_{2}}\delta_{\lambda_{1}+2\mu_{1}+\lambda_{2}+2\mu_{2},m_{1}+m_{2}+m_{3}}
\times$$

$$\left[{(\lambda_{2}+1)(\lambda_{2}^{\prime}+\mu_{2}^{\prime}-\mu_{2})!
\mu_{2}!(m_{1}-m_{2}+1)(m_{1}-m_{3}+2)(m_{2}-m_{3}+1)
(m_{12}^{\prime\prime}-m_{2})!
\over{(\lambda_{2}+\mu_{2}-\lambda_{2}^{\prime}-\mu_{2}^{\prime})!(\mu_{2}-\mu_{2}^{\prime})!
(\lambda_{2}+\mu_{2}-m_{12}^{\prime})!
(m_{2}-k_{2}+1)!(m_{1}-m_{12}^{\prime\prime})!(m_{2}-m_{22}^{\prime\prime})!}}
\right.\times$$

$${(\lambda-k_{1}-k_{2}-m_{2})!(\mu^{\prime}_{1}+k_{1}-m_{3})!(\lambda-k_{1}-k_{2}-m_{3}+1)!(m_{22}^{\prime\prime}-m_{3})!
(m_{12}^{\prime\prime}-m_{3}+1)!\over{
(m_{1}-\lambda+k_{1}+k_{2})!(m_{2}-\mu_{1}-k_{1})!(m_{3}-k_{2})!(m_{1}-\mu_{1}-k_{1}+1)!(m_{1}-k_{2}+2)!}}
\times$$

$$\left.{(m_{12}^{\prime}-\mu_{2})!(\mu_{1}-k_{2})!(\lambda_{1}+\mu_{1}-k_{2}+1)!(\lambda_{1}+\mu_{1}-m_{12})!(\mu_{1}-m_{22})!
(\lambda_{1}+\mu_{1}-m_{22}+1)!\over{(\mu_{2}-m_{22}^{\prime})!(\lambda_{2}+\mu_{2}-m_{22}^{\prime}+1)!
(\mu_{1}+k_{1}+1)!(m_{12}-\mu_{1})!m_{22}!(m_{12}+1)!(m_{1}-m_{22}^{\prime\prime}+1)!}}\right]^{1/2}\times$$

$$\sum^{\mu^{\prime}_{2}}_{p=0}~\sum^{\max(q)}_{q=\min(q)}(\mu_{2}^{\prime}-p)!
(\lambda^{\prime}_{2}+\mu^{\prime}_{2}+p-m_{12}^{\prime}
-m_{22}^{\prime})!\left({(m_{12}^{\prime\prime}+m_{22}^{\prime\prime}-p-q-m_{12})!(p-m_{22}^{\prime})!(q-m_{22})!
\over{(m_{12}^{\prime}-p)!}}\right)^{1/2}\times$$

$$\left({(m_{12}^{\prime\prime}
+m_{22}^{\prime\prime}-p-q-m_{22}+1)!
(m_{12}^{\prime\prime}+m_{22}^{\prime\prime}-p-2q+1)(p+q-m_{22}^{\prime\prime})!
(m_{22}^{\prime\prime}-q+1)!\over{(m_{12}^{\prime\prime}-p-q)!(m_{12}^{\prime\prime}
+m_{22}^{\prime\prime}-p-q-\mu_{1}-k_{1})!(m_{12}-q)!}}\right) ^{1/2}\times$$

$$U\left(\matrix{[m_{12}m_{22}]
&[m_{12}^{\prime}+m_{22}^{\prime}-p~0] &[m_{12}^{\prime\prime}+m_{22}^{\prime\prime}-p-q,q]\cr
[p0] &[m_{12}^{\prime\prime}~m_{22}^{\prime\prime}] &[m_{12}^{\prime}~m_{22}^{\prime}]\cr}\right)\times$$

$$F_{2}\left(\matrix{[\lambda-k_{1}-k_{2},\mu_{1}+k_{1},k_{2}]
&[m_{1}m_{2}m_{3}]\cr
[m_{12}^{\prime\prime}+m_{22}^{\prime\prime}-p-q,q]
&[m_{12}^{\prime\prime}~m_{22}^{\prime\prime}]\cr}\right)\times$$

$$F_{2}\left(\matrix{[\lambda_{1}+\mu_{1}~\mu_{1}]
 &[\lambda-k_{1}-k_{2},\mu_{1}+k_{1},k_{2}]\cr
[m_{12}+m_{22}~0]
&[m_{12}^{\prime\prime}+m_{22}^{\prime\prime}-p-q,q]\cr}\right)\times$$

$$F\left(\matrix{\lambda_{2}^{\prime}+\mu_{2}^{\prime}\cr m^{\prime}_{12}+m_{22}^{\prime}-p\cr};
\matrix{\mu_{2}^{\prime}\cr p\cr};\matrix{[\lambda_{2}+\mu_{2}~\mu_{2}]\cr
[m_{12}^{\prime}~m_{22}^{\prime}]\cr}\right),\eqno(5.7)$$
\vskip .3cm
\noindent where we have written explicitly that the irrep $(\lambda_{2}\mu_{2})$ is coupled from
$(\lambda_{2}^{\prime}0)\times (\mu_{2}^{\prime}0)$ with $\lambda_{2}^{\prime}+2\mu_{2}^{\prime}=
\lambda_{2}+2\mu_{2}$, 
\vskip .3cm
$$\lambda=\lambda_{1}+\mu_{1}+\lambda^{\prime}_{2}+\mu^{\prime}_{2},$$

$$[u_{1},u_{2},u_{3}]=[\lambda-k_{1}-k_{2},\mu_{1}+k_{1},k_{2}],\eqno(5.8)$$
\vskip .3cm
\noindent and 
\vskip .3cm
$$F\left(\matrix{\lambda_{2}^{\prime}+\mu_{2}^{\prime}\cr m^{\prime}_{12}+m_{22}^{\prime}-p\cr};
\matrix{\mu_{2}^{\prime}\cr p\cr};\matrix{[\lambda_{2}+\mu_{2}~\mu_{2}]\cr
[m_{12}^{\prime}~m_{22}^{\prime}]\cr}\right)=$$

$$\sum_{t}(-)^{t}
{(m_{12}^{\prime}-p+t)!(\lambda_{2}+\mu_{2}-m_{12}^{\prime}-m_{22}^{\prime}+p-t)!\over{
t!(\lambda_{2}^{\prime}+\mu_{2}^{\prime}-m_{12}^{\prime}-m_{22}^{\prime}+p-t)!(p-m_{22}^{\prime}-t)!
(m_{12}^{\prime}+m_{22}^{\prime}-p-\mu_{2}+t)_!}}.\eqno(5.9)$$

\vskip .6cm
\begin{centering}
{\large IV. Discussions }\\
\end{centering}
\vskip .4cm
   In this paper, based on the complementary group technique for the resolution of outer multiplicity
problem of $SU(n)$ proposed in (I), a general procedure for the derivation of $SU(3)\supset U(2)$ RWCs with
multiplicity is outlined by using the recoupling approach after a special transformation. It is proved that 
outer multiplicity labeling scheme is not unique, and can be transformed from one another within  $SO(m)$, where
$m$ is the multiplicity of $[m_{1}m_{2}m_{3}]$ in the decomposition
$(\lambda_{1}\mu_{1})\times(\lambda_{2}\mu_{2})$. From this point of view, Biedenharn's resolution for the outer
multiplicity of $SU(n)$ can also be regarded as a complementary group resolution. In their works, the
complementary group is $U(n)$. Unlike the resolution proposed in (I), in which only special Gel'fand basis was
considered according to the Littlewood rule, the subirreps of the complementary group $U(n)$ will all be
considered in order to resolve the multiplicity of $SU(n)\times SU(n)$. That is why multiplicity formulae can not
easily be worked out from their outer multiplicity labeling scheme. Actually, all these canonical resolutions are
equivalent, and can be transformed from each other. Therefore, all canonical resolutions form an equivalent
class. 
\vskip .3cm
   Using this method, RWCs of $SU(3)\supset U(2)$ with multiplicity can be derived recursively. It is obvious
that computer code  based on this procedure can easily be compiled, which will make numerical calculation
possible in practical applications. Furthermore, one can also obtain closed  algebraic expressions of these
RWCs for small $m$ values. However, the expression is still cumbersome with summation over many variables.
The complexity increases with increasing the multiplicity. Therefore, a further simplification is till expected.
\vskip .3cm
   Only the RWCs of $SU(3)$ in the canonical chain $SU(3)\supset U(2)$ are discussed. In fact, this method
can also be applied to noncanonical basis of $SU(3)$, for example $SU(3)\supset SO(3)$, by using RWCs of
$SU(3)\supset SO(3)$ with one irrep symmetric in the coupling, which is multiplicity-free, and may be easily 
obtained.$^{[40-41]}$ In addition, the procedure outlined in this paper can also be extended to general $SU(n)$
case. We will discuss  $SU(4)\supset U(3)$  RWCs in the next paper.
\vskip .3cm
   A closed algebraic expression of reduced AWCs proposed in [11] is also obtained by using the recoupling
approach. These coefficients satisfy  different orthogonality relations, and may be useful in many-particle
coupling problems.
\vskip .5cm
\noindent {\bf Acknowledgment}
\vskip .5cm
   The project was supported by US National Science Foundation and partly by the State Education
Commission of China.
   
\vskip .5cm
\begin{tabbing}
\=1111\=22222222222222222222222222222222222222222222222222222222222222222222222222222222\=\kill\\
\>{[1]}\>{G. E. Baird, and L. C. Biedenharn, J. Math. Phys., {\bf 4}(1963) 1449; {\bf 5}(1964) 1723;}\\
\>{}\>{{\bf 5}(1964) 1730; {\bf 6}(1965) 1847}\\
\>{[2]}\>{L. C. Biedenharn, A. Giovannini, and J. D. Louck, J. Math. Phys., {\bf 8}(1967) 691}\\
\>{[3]}\>{L. C. Biedenharn, J. D. Louck, Commun. Math. Phys., {\bf 8}(1968) 89}\\
\>{[4]}\>{L. C. Biedenharn, J. D. Louck, E. Chacon, and M.  Ciftan, J. Math. Phys., {\bf 13}(1972)
1957}\\
\>{[5]}\>{E. Chacon, M. Ciftan, and L. C. Biedenharn, J. Math. Phys., {\bf 13}(1972) 577}\\
\>{[6]}\>{L. C. Biedenharn, and J. D. Louck, Commun. Math. Phys., {\bf 93}(1984) 143}\\
\>{[7]}\>{L. C. Biedenharn, M. A. Lohe, and J. D Louck, J. Math. Phys., {\bf 26} (1985) 1458}\\
\>{[8]}\>{R. Le Blanc and L. C. Biedenharn, J. Phys. A, {\bf 22}(1989) 4613}\\
\>{[9]}\>{K. T. Hecht and L. C. Biedenharn, J. Math. Phys., {\bf 31}(1990) 2781}\\
\>{[10]}\>{L. C. Bidenharn, M. A. Lohe, and H. T. Williams, J. Math. Phys., {\bf 35}(1994) 6072}\\
\>{[11]}\>{T. A. Brody, M. Moshinsky, and I. Renero, J. Math. Phys., {\bf 6}(1965) 1540}\\
\>{[12]}\>{M. Moshinsky, J. Math. Phys., {\bf 4}(1963) 1128}\\
\>{[13]}\>{M. Moshisky, Rev. Mod. Phys., {\bf 34}(1962) 813}\\
\>{[14]}\>{M. Moshinsky, J. Math. Phys., {\bf 7}(1966) 691}\\
\>{[15]}\>{S. J. Ali\v{s}auskas, A. A. Jucy, and A. P. Jucy, J. Math. Phys., {\bf 13}(1972) 1329}\\
\>{[16]}\>{S. J. Ali\v{s}uaskas, V. V. Vanagas, and J. P. Jucy, Dokl. Akad. Nauk. SSSR, {\bf 197}(1971) 804}\\
\>{[17]}\>{S. J. Ali\v{s}uaskas, Sov. J. Part. Nucl., {\bf 14} (1983) 563}\\
\>{[18]}\>{S. J. Ali\v{s}auskas, J. Math. Phys., {\bf 29} (1988) 2351}\\
\>{[19]}\>{S. J. Ali\v{s}auskas, J. Math. Phys., {\bf 31} (1990) 1325}\\
\>{[20]}\>{S. J. Ali\v{s}auskas, J. Math. Phys., {\bf 33} (1992) 1983}\\
\>{[21]}\>{S. J. Ali\v{s}auskas, J. Phys. A, {\bf 29} (1996) 2687}\\
\>{[22]}\>{K. T. Hecht, Nucl Phys, {\bf 62} (1965) 1}\\
\>{[23]}\>{M. Resnikoff, J. Math. Phys., {\bf 8} (1967) 63}\\
\>{[24]}\>{L. A. Shelepin and V. P. Karasev, Sov. J. Nucl. Phys., {\bf 5} (1967) 156}\\
\>{[25]}\>{V. P. Karasev and L. A. Shelepin, Sov. J. Nucl. Phys., {\bf 7} (1968) 678}\\
\>{[26]}\>{A. U. Klimyk and A. M. Gavrilik, J. Math. Phys., {\bf 20}(1979) 1624}\\
\>{[27]}\>{R. Le Blanc and D. J. Rowe, J. Phys. A, {\bf 19} (1986) 2913}\\
\>{[28]}\>{J. P. Draayer and Y. Akiyama, J. Math. Phys. {\bf 14} (1973) 1904}\\
\>{[29]}\>{Y. Akiyama and J. P. Draayer, Comp. Phys. Commun., {\bf 5} (1973) 405}\\
\>{[30]}\>{T. A. Kaeding, Comp. Phys. Commun., {\bf 85} (1995) 82}\\
\>{[31]}\>{T. A. Kaeding and H. T. Williams, Commp. Phys. Commun., {\bf 98} (1996) 398}\\
\>{[32]}\>{H. T. Williams, J. Math. Phys., {\bf 37} (1996) 4187}\\
\>{[33]}\>{J. A. Castilho  Alcaras, L. C. Biedenharn, K. T. Hecht, and G. Neely, Ann. Phys., {\bf 60}(1970) 85}\\
\>{[34]}\>{J. S. Prakash and H. S. Sharatchandra, J. Math. Phys., {\bf 37} (1996) 6530}\\
\>{[35]}\>{D. Braunschweig, Comp. Phys. Commun., {\bf 14}(1978) 109}\\
\>{[36]}\>{K. T. Hecht, Nucl. Phys. {\bf 62}(1965) 1; {\bf 63}(1965) 177; {\bf A102}(1967) 11}\\
\>{[37]}\>{K. T. Hecht and S. C. Pang, J. Math. Phys., {\bf 10}(1969) 1571}\\
\>{[38]}\>{D. Millener, J. Math. Phys., {\bf 19}(1978) 1513}\\
\>{[39]}\>{J. J. De Swart, Rev. Mod. Phys., {\bf 35}(1963) 916}\\
\>{[40]}\>{J. D. Vargados, Nucl. Phys., {\bf A111}(1968) 681}\\
\>{[41]}\>{S. J. Ali\v{s}auskas,  E. Norvaisas, J. Phys., {\bf A22}(1989)5177}\\
\end{tabbing}
\newpage

{\bf Table I.}~ RWCs of $SU(3)\supset U(2)$ $\left<\matrix{[21] &[21]\cr
[q_{1}q_{2}]
&[q_{1}^{\prime}q_{2}^{\prime}]\cr}\right\vert\left.\matrix{\eta~[m_{1}m_{2}m_{3}]\cr
[32]\cr}\right>$\\
\begin{tabbing}
\=11111111111111111111111111111111111111111111111111111111111111111111111111111111111111111111111111111111\=\kill\\
{---------------------------------------------------------------------------------------------------------------}\\
\=1111111111111111111111111\=2222222222\=33333333333\=444444444444\=555555555555\=\kill\\
\>{$\downarrow~[q_{1}q_{2}]~[q^{\prime}_{1}q_{2}^{\prime}]~/~[m_{1}m_{2}m_{3}]$}\>{$[420]$}\>{$[321]_{1}$}\>{$[321]_{2}$}\>{$[300]$}\\
\=11111111111111111111111111111111111111111111111111111111111111111111111111111111111111111111111111111111\=\kill\\
\>{---------------------------------------------------------------------------------------------------------------}\\
\=1111111111111111111111111\=2222222222\=33333333333\=444444444444\=555555555555\=\kill\\
\>{~~[21]~~[20]}\>{$-\sqrt{1\over{20}}$}\>{$-\sqrt{9\over{20}}$}\>{$\sqrt{1\over{4}}$}\>{$\sqrt{1\over{4}}$}\\
\>{}\\
\>{~~[20]~~[21]}\>{$\sqrt{1\over{20}}$}\>{$\sqrt{9\over{20}}$}\>{$\sqrt{1\over{4}}$}\>{$\sqrt{1\over{4}}$}\\
\>{}\\
\>{~~[21]~~[11]}\>{$\sqrt{9\over{20}}$}\>{$-\sqrt{1\over{20}}$}\>{$-\sqrt{1\over{4}}$}\>{$\sqrt{1\over{4}}$}\\
\>{}\\
\>{~~[11]~~[21]}\>{$\sqrt{9\over{20}}$}\>{$-\sqrt{1\over{20}}$}\>{$\sqrt{1\over{4}}$}\>{$-\sqrt{1\over{4}}$}\\
\=11111111111111111111111111111111111111111111111111111111111111111111111111111111111111111111111111111111\=\kill\\
{---------------------------------------------------------------------------------------------------------------}\\
{---------------------------------------------------------------------------------------------------------------}\\
\=1111111111111111111111111\=2222222222\=33333333333\=444444444444\=555555555555\=\kill\\
\>{~~[21]~~[20]}\>{$-\sqrt{1\over{20}}$}\>{$\sqrt{7\over{10}}$}\>{$0$}\>{$\sqrt{1\over{4}}$}\\
\>{}\\
\>{~~[20]~~[21]}\>{$\sqrt{1\over{20}}$}\>{$-\sqrt{2\over{35}}$}\>{$\sqrt{9\over{14}}$}\>{$\sqrt{1\over{4}}$}\\
\>{}\\
\>{~~[21]~~[11]}\>{$\sqrt{9\over{20}}$}\>{$-\sqrt{1\over{70}}$}\>{$-\sqrt{2\over{7}}$}\>{$\sqrt{1\over{4}}$}\\
\>{}\\
\>{~~[11]~~[21]}\>{$\sqrt{9\over{20}}$}\>{$\sqrt{8\over{35}}$}\>{$\sqrt{1\over{14}}$}\>{$-\sqrt{1\over{4}}$}\\
\=11111111111111111111111111111111111111111111111111111111111111111111111111111111111111111111111111111111\=\kill\\
\>{---------------------------------------------------------------------------------------------------------------}\\
\=1111111111111111111111111\=2222222222\=33333333333\=444444444444\=555555555555\=\kill\\
\>{$\uparrow~[q_{1}q_{2}]~[q^{\prime}_{1}q_{2}^{\prime}]~/~[m_{1}m_{2}m_{3}]$}\>{$[420]$}\>{$[321]_{\eta=0}$}\>{$[321]_{\eta=1}$}\>{$[300]$}\\
\=11111111111111111111111111111111111111111111111111111111111111111111111111111111111111111111111111111111\=\kill\\
\>{---------------------------------------------------------------------------------------------------------------}\\
\end{tabbing}

\newpage

{\bf Table II.}~ RWCs of $SU(3)\supset U(2)$ $\left<\matrix{[21] &[21]\cr
[q_{1}q_{2}]
&[q_{1}^{\prime}q_{2}^{\prime}]\cr}\right\vert\left.\matrix{\eta~[m_{1}m_{2}m_{3}]\cr
[22]\cr}\right>$\\
\begin{tabbing}
\=11111111111111111111111111111111111111111111111111111111111111111111111111111111111111111111111111111111\=\kill\\
{---------------------------------------------------------------------------------------------------------------}\\
\=1111111111111111111111111\=2222222222\=33333333333\=444444444444\=555555555555\=\kill\\
\>{$\downarrow~[q_{1}q_{2}]~[q^{\prime}_{1}q_{2}^{\prime}]~/~[m_{1}m_{2}m_{3}]$}\>{$[420]$}\>{$[321]_{1}$}\>{$[321]_{2}$}\>{$[330]$}\\
\=11111111111111111111111111111111111111111111111111111111111111111111111111111111111111111111111111111111\=\kill\\
\>{---------------------------------------------------------------------------------------------------------------}\\
\=1111111111111111111111111\=2222222222\=33333333333\=444444444444\=555555555555\=\kill\\
\>{~~[10]~~[21]}\>{$\sqrt{3\over{20}}$}\>{$\sqrt{1\over{10}}$}\>{$\sqrt{1\over{2}}$}\>{$\sqrt{1\over{4}}$}\\
\>{}\\
\>{~~[21]~~[10]}\>{$-\sqrt{3\over{20}}$}\>{$-\sqrt{1\over{10}}$}\>{$\sqrt{1\over{2}}$}\>{$-\sqrt{1\over{4}}$}\\
\>{}\\
\>{~~[20]~~[20]}\>{$-\sqrt{1\over{40}}$}\>{$-\sqrt{3\over{5}}$}\>{$0$}\>{$\sqrt{3\over{8}}$}\\
\>{}\\
\>{~~[11]~~[11]}\>{$\sqrt{27\over{40}}$}\>{$-\sqrt{1\over{5}}$}\>{$0$}\>{$-\sqrt{1\over{8}}$}\\
\=11111111111111111111111111111111111111111111111111111111111111111111111111111111111111111111111111111111\=\kill\\
\>{---------------------------------------------------------------------------------------------------------------}\\
\>{---------------------------------------------------------------------------------------------------------------}\\
\=1111111111111111111111111\=2222222222\=33333333333\=444444444444\=555555555555\=\kill\\
\>{~~[10]~~[21]}\>{$\sqrt{3\over{20}}$}\>{$\sqrt{16\over{35}}$}\>{$-\sqrt{1\over{7}}$}\>{$\sqrt{1\over{4}}$}\\
\>{}\\
\>{~~[21]~~[10]}\>{$-\sqrt{3\over{20}}$}\>{$\sqrt{1\over{35}}$}\>{$-\sqrt{4\over{7}}$}\>{$-\sqrt{1\over{4}}$}\\
\>{}\\
\>{~~[20]~~[20]}\>{$-\sqrt{1\over{40}}$}\>{$-\sqrt{27\over{70}}$}\>{$-\sqrt{3\over{14}}$}\>{$\sqrt{3\over{8}}$}\\
\>{}\\
\>{~~[11]~~[11]}\>{$\sqrt{27\over{40}}$}\>{$-\sqrt{9\over{70}}$}\>{$-\sqrt{1\over{14}}$}\>{$-\sqrt{3\over{8}}$}\\
\=11111111111111111111111111111111111111111111111111111111111111111111111111111111111111111111111111111111\=\kill\\
\>{---------------------------------------------------------------------------------------------------------------}\\
\=1111111111111111111111111\=2222222222\=33333333333\=444444444444\=555555555555\=\kill\\
\>{$\uparrow~[q_{1}q_{2}]~[q^{\prime}_{1}q_{2}^{\prime}]~/~[m_{1}m_{2}m_{3}]$}\>{$[420]$}\>{$[321]_{\eta=0}$}\>{$[321]_{\eta=1}$}\>{$[330]$}\\
\=11111111111111111111111111111111111111111111111111111111111111111111111111111111111111111111111111111111\=\kill\\
\>{---------------------------------------------------------------------------------------------------------------}\\
\end{tabbing}
\end{document}